\documentclass[twocolumn]{aa}  
\usepackage[varg]{txfonts}
\usepackage[utf8]{inputenc}
\usepackage[english]{babel}
\usepackage{natbib}
\usepackage{xcolor}
\usepackage{url}
\bibpunct{(}{)}{;}{a}{}{,}
\usepackage{multirow}
\usepackage{enumerate}
\usepackage[inline]{enumitem}

\usepackage{amstext}            
\usepackage{array}              
\newcolumntype{L}{>{$}l<{$}}    
\newcolumntype{R}{>{$}r<{$}}
\newcolumntype{C}{>{$}c<{$}}
\usepackage{caption} 
\usepackage{subcaption}
\usepackage[singlelinecheck=false]{caption} 
\usepackage{booktabs} 
\usepackage{graphicx}
\usepackage{txfonts}
\usepackage{hyperref}
\hypersetup{colorlinks=True, citecolor={blue}, urlcolor={blue},linkcolor={black}}
\usepackage[capitalise]{cleveref}
\usepackage{placeins}
\DeclareUnicodeCharacter{2212}{-}
\newcommand{\XMM}{\textit{XMM-Newton}\xspace}
\newcommand{\CHANDRA}{\textit{Chandra}\xspace}

\begin{document} 

   \title{LOFAR detection of extended emission around a mini-halo \\in the galaxy cluster Abell 1413}

   \author{G. Lusetti\inst{1,2}
          \and
          A. Bonafede\inst{2}
          \and
          L. Lovisari\inst{7,8}
          \and
          M. Gitti\inst{2}
          \and
          S. Ettori\inst{3}
          \and
          R. Cassano\inst{4}
          \and
          C. J. Riseley\inst{2,4,5}
          \and \\
          F. Govoni\inst{12}
          \and
          M. Br\"uggen\inst{1}
          \and
          L. Bruno\inst{2,4}
          \and
          R. J. van Weeren\inst{6}
          \and
          A. Botteon\inst{4}
          \and 
          D. N. Hoang\inst{1}
          \and \\
          F. Gastaldello\inst{9}
          \and
          A. Ignesti\inst{10}
          \and
          M. Rossetti\inst{9}
          \and
          T. W. Shimwell\inst{11}
          }

   \institute{Hamburger Sternwarte, University of Hamburg, Gojenbergsweg 112, 21029 Hamburg, Germany \\
            \email{giulia.lusetti@hs.uni-hamburg.de}
    \and 
             DIFA - Università di Bologna, via Gobetti 93/2, I-40129 Bologna, Italy
    \and 
            INAF - Osservatorio di Astrofisica e Scienza dello Spazio di Bologna, via Piero Gobetti 93/3, 40129 Bologna, Italy 
    \and 
            INAF - Istituto di Radioastronomia di Bologna, Via Gobetti 101, 40129 Bologna, Italy
    \and 
            CSIRO Space \& Astronomy, PO Box 1130, Bentley, WA 6102, Australia
    \and 
            Leiden Observatory, Leiden University, PO Box 9513, 2300 RA Leiden, The Netherlands
    \and 
            INAF, Istituto di Astrofisica Spaziale e Fisica Cosmica di Milano, via A. Corti 12, 20133 Milano, Italy
    \and 
            Center for Astrophysics $|$ Harvard $\&$ Smithsonian, 60 Garden Street, Cambridge, MA 02138, USA
    \and 
            INAF - IASF Milano, Via A. Corti 12, I-20133, Milano, Italy
    \and 
            INAF - Padova Astronomical Observatory, Vicolo dell'Osservatorio 5, I-35122 Padova, Italy
    \and 
            ASTRON, Netherlands Institute for Radio Astronomy, Oude Hoogeveensedijk 4, 7991 PD, Dwingeloo, The Netherlands and Leiden Observatory, Leiden University, PO Box 9513, NL-2300 RA Leiden, The Netherlands
    \and 
            INAF - Osservatorio Astronomico di Cagliari, Via della Scienza 5, I-09047 Selargius (CA), Italy
             }

   \date{August 2023}

  \abstract
   {The relation between giant radio halos and mini-halos in galaxy clusters is not understood. The former are usually associated with merging clusters, while the latter are found in relaxed systems. In the last years, the advent of low-frequency radio observations has challenged this dichotomy by finding intermediate objects with a hybrid radio morphology.}
   {We investigate the presence of diffuse radio emission in the cluster Abell 1413 and determine its dynamical status to explore the relation between mini-halos and giant radio halos.}
   {We use LOFAR observations centered at 144 MHz to study the diffuse radio emission. To investigate the dynamical state of the system, we use newly analysed \XMM archival data.}
   {
   A1413 shows features that are typically present in both relaxed (e.g., peaked x-ray surface brightness distribution and little large scale inhomogeneities) and disturbed  (e.g., flatter temperature and metallicity profiles) clusters.
   This suggests that A1413 is neither disturbed nor fully relaxed, and we argue that it is an intermediate-phase cluster. 
   At 144 MHz, we discover a wider diffuse component surrounding the previously known mini-halo at the cluster center. By fitting the radio surface brightness profile with a double-exponential model, we can disentangle the two components. We find an inner mini-halo with an $e$-folding radius $r_{\rm e,1} = 28\pm 5\,\rm kpc$ and an extended component with $r_{\rm e,2} = 290\pm60\,\rm kpc$. We also evaluated the point-to-point correlation between the radio and X-ray surface brightness, finding a sub-linear relation for the outer emission and a super-linear relation for the mini-halo.
   The mini-halo and the diffuse emission extend over different scales and show different features, confirming the double nature of the radio emission and suggesting that the mechanisms responsible for the re-acceleration of the radio emitting particle might be different.
   }
   {}

   \titlerunning{Extended emission around the mini-halo in A1413}
   \authorrunning{Lusetti et al.}
   \maketitle 


\section{Introduction}
\label{sec:Intro}
Galaxy clusters have a size that ranges from a few hundreds kpc to $\sim$ Mpc, and have typical masses of $10^{14}-10^{15}\, \mathrm{M_\odot}$. They can contain up to $\sim1000$ galaxies, embedded in the intracluster medium (ICM). The ICM consists of fully ionized hydrogen and helium plus traces of highly ionized heavier elements with a temperature $ \mathrm{T} \sim 10^{7}-10^8\, \mathrm{K} \,(1-10\, \mathrm{keV})$ and particle number densities steeply declining from $n_e \sim 10^{-2}-10^{-3}\, \mathrm{cm^{−3}}$ near the center to $10^{-4} \mathrm{cm^{−3}}$ in the outskirts. It emits mainly via thermal bremsstrahlung in the X-ray band \cite[e.g.,][]{FormanJones1982, Sarazin2009}.

Co-spatial with the ICM, an increasing number of galaxy clusters is also observed to host diffuse non-thermal synchrotron emission, with no obvious connection to the individual cluster members. The synchrotron emission reveals the presence of magnetic field and relativistic cosmic-ray electrons (CRe) in the ICM. Extended, central diffuse sources are usually classified as giant radio halos and mini-halos \cite[see][for a review]{vanWeeren2019}. 
Both giant and mini-halos exhibit steep spectrum\footnote{The synchrotron radio spectrum is defined as $S(\nu)\propto\nu^{-\alpha}$.} with a spectral index $\rm\alpha>1$. 
While giant radio halos are Mpc-sized radio sources which cover large parts of the cluster volume, mini-halos are  typically confined within the cool-cores ( $\sim 300-500$ kpc) of relaxed galaxy clusters, and are characterized by higher emissivity \cite[e.g.,][]{Cassano2008, Murgia2009}.

Two emission mechanisms have been proposed to explain these non-thermal radio sources \cite[see][for a review]{Brunetti&Jones2014}: ($i$) the (re-)acceleration scenario, where CRe are (re-)accelerated by turbulence that develops during cluster mergers; ($ii$) the hadronic scenario, where secondary CRe are produced by proton-proton collisions between relativistic cosmic-ray protons (CRp) and thermal protons in the ICM. 
While it has been shown that purely hadronic models cannot reproduce the emission of giant radio halos \citep{Brunetti2008, Brunetti2017}, they remain a viable mechanism for mini-halos \citep{Pfrommer&Ensslin2004, JacobandPfrommer2017}.
%
Statistical studies have shown the connection between the dynamical status of the cluster and the type of diffuse radio emission \cite[e.g.,][]{Cassano2010, Cuciti2021, Cassano2023}. 
Galaxy clusters hosting giant halos are mainly characterized by a disturbed dynamical status, suggesting recent or ongoing mergers \cite[e.g.,][]{Cassano2013, Cuciti2015}. 
On the other hand, mini-halos are usually found in relaxed cool-core clusters \citep{Gitti2004, Cassano2010, Gitti2018} but their origin is unclear since they could be of leptonic or hadronic origin \cite[e.g.,][]{vanWeeren2019}. 
Moreover, in a statistical study about the occurrence of mini-halos in a sample of massive ($\rm M>6\cdot 10^{14}M_{\odot}$) galaxy clusters, \cite{Giacintucci2017} found that $\rm \sim 80\%$ of cool-core clusters (classified according to the value of central entropy $\rm K_{0}<30\,keV\, cm^2$) host a mini-halo, suggesting a clear relation between the origin of mini-halos and the properties of the cool-core gas.

In the last years, the advent of deep, low-frequency observations has complicated the picture described above. In fact, a few cool-core clusters show the presence of a diffuse synchrotron halo or halo-like emission that extends quite far from the central mini-halo \citep{Bonafede2014, Venturi2017, Sommer2017, Savini2018, Savini2019, Biava2021, Riseley2022b, Bruno2023_A2142}, challenging the idea that  mini-halos occur in cool-core clusters and giant halos in non cool-core clusters. 
These peculiar galaxy clusters, either, present giant halos \citep{Bonafede2014, Savini2018}, or show a hybrid morphology in the radio emission \citep{Biava2021}, even though being classified as relaxed systems based on their thermal properties. Until now, a somewhat unique case is represented by the galaxy cluster RX J1720.1+2638 \citep{Savini2019, Biava2021}, where the radio emission is composed of a central bright mini-halo plus fainter and wider emission, extending far from the cluster core. In this study, the authors showed that the radio diffuse emission presents different features, suggesting an intrinsic different nature of the two sources.
An even more complex morphology has been found in Abell 2142, a galaxy cluster with an intermediate dynamical state, already known to host a hybrid radio halo with two distinct components \citep{Venturi2017}. This object has recently been studied with low-frequency radio observation by \cite{Bruno2023_A2142}, who detected a third ultra-steep spectrum and wider component.
Only a handful of multi-component diffuse sources only are currently known, but their number may increase thanks to sensitive low frequency observations. \\

%
In this work we present the galaxy cluster Abell 1413 (hereafter A1413; Planck name: PSZ2 G226.18+76.79) is a nearby (z=0.143) massive ($M_{500}=5.95^{+0.24}_{-0.25}\times 10^{14} \, M_\odot$, \citealt{PlanckCollaboration2016}) system. It is well-studied in the optical \citep{Castagne2012}, X-ray \citep{Pratt2002} and radio \citep{Govoni2001, Giacintucci2017, Savini2019} bands.
These previous studies at different wavelenghts showed that the properties of A1413 are intermediate between classical relaxed and merging clusters.
From the morphological point of view, it shows a high ellipticity confirmed by both optical \citep{Castagne2012} and X-ray \citep{Pratt2002} studies. The cluster is elongated approximately in the North-South direction (Fig. 7 of \citealt{Castagne2012}). 
Studying the galaxy number density distribution, \cite{Castagne2012} estimated the value of the cluster ellipticity $\epsilon \sim 0.35$ at large radii, increasing up to $\sim0.8$ at the cluster center $r < 1\arcmin$. While fitting the \XMM image, they found values quickly converging toward $\epsilon \sim 0.27$, which is fully compatible with the global value of $\epsilon \sim 0.29$ derived by \cite{Pratt2002} over the radial range $[3\arcmin - 13\arcmin]$.
They argue that this discrepancy at large radii of the optical and X-ray ellipticity values is not surprising as the collisional gas relaxes more quickly within the cluster potential than the non-collisional galaxies and reaches more rapidly a spherically symmetric distribution. However, \cite{Castagne2012} also stressed that A1413 seems a rather extreme example of this effect.
Moreover, by analysing the velocity distribution of the cluster, they found a velocity offset of $450\pm210\,\rm km\,s^{-1} $(i.e., $2\sigma$ significance) between the velocity of the central cD galaxy \citep{Paturel1989}, also being the brightest cluster galaxies (BCG), and the mean cluster velocity. These authors argue that BCGs velocity offsets are frequently found in clusters that are distant from equilibrium, for example during significant merging events. If this BCG velocity offset is confirmed, it would indicate that A1413 may  potentially be in a non-fully relaxed dynamical state.

On the other hand, early X-ray studies \citep{Pratt2002, Vikhlinin2005, Baldi2007} classified this cluster as a relaxed system, with no obvious signatures of a recent merger. The lack of shocks or cold fronts has been also confirmed by \cite{Botteon2018}, who explored a combination of different analysis approaches of X-ray observations to firmly detect and characterize edges in a sample of non-cool-core,
massive galaxy clusters.
The \XMM observation \citep{Pratt2002} does not provide strong evidence of a cool-core region, in contrast with the temperature profiles obtained with \CHANDRA \citep{Vikhlinin2005}, with a cooling time in the innermost bin ($\rm\sim 50\, kpc$) of $t_{\rm cool}=4.2\pm0.3$ Gyrs \citep{Baldi2007}.
These authors suggested that this could be related to the better resolution of \CHANDRA compared to \XMM.
Finally, using common X-ray morphological parameters, such as the concentration parameter ($c$, \citet{Santos2008}) and the centroid shift ($w$,  \citet{Poole2006, Mohr1995}) A1413 has been classified as a relaxed system \citep{Lovisari2017, Campitiello2022, Yuan2022}. However, it also has a mixed morphology, i.e. elongated X-ray isophotes and a relatively flat X-ray profile near the centre.

The first investigation of the radio emission in A1413 was performed by \cite{Govoni2009}, who found a candidate mini-halo at a frequency of 1.4 GHz.  
Later, A1413 was included in a mini-halos statistical study of \cite{Giacintucci2017}, who classified clusters according to the value of the specific central entropy $K_0$. 
Interestingly, A1413 was the only non-cool-core cluster ($K_0 = 64 \pm 8\,\rm keV cm^2$) of the whole sample hosting a (candidate) mini-halo.
The diffuse radio emission was then studied at a frequency of 144 MHz, in a study of non-merging galaxy clusters \citep{Savini2019}, confirming the presence of a mini-halo with a linear size of $\sim$ 210 kpc and a total flux density of $40\pm7$ mJy.
In a recent study of \cite{Riseley2023}, the mini-halo was studied with a combination of LOFAR at 144 MHz and MeerKAT at 1283 MHz observations. They measured a linear size of at least 449~kpc (and up to 584~kpc) at 1283 MHz, twice as large as previously measured by \cite{Govoni2009}.
From these past studies, a controversial and intriguing picture of A1413 emerges, making it a very interesting object. In this paper we present new radio observations performed with the LOFAR High Band Antennae (HBA) at 144 MHz. This is combined with archival and newly analysed \XMM data, which includes a more accurate background treatment and improved calibration. 

This paper is organised as follows: in \cref{sec:datared} we give an overview of the observations and data reduction. The X-ray and radio results are presented in \cref{sec:results}. 
We discuss the results from X-ray and radio analysis jointly in \cref{sec:discussion}. In \cref{sec:conclusion} we summarise our conclusions.

Throughout the paper we adopt a $\Lambda$CDM cosmology with $\mathrm{H_0 = 70 \,km\,s^{-1} Mpc^{-1}}$, $\Omega_\Lambda = 0.7$ and $\Omega_{\rm m} = 0.3$. Thus, $1''$ corresponds to a physical scale of $2.51$ kpc at the redshift of A1413. A summary of the properties of A1413 collected from literature is reported in \cref{tab:properties}.

\begin{table}
    \caption{Properties of A1413 derived from literature: redshift (z: \citealt{Boehringer2000}), equatorial coordinates (NASA ExtragalCatalog, \citealt{Bade1998}), mass within the radius that defines the sphere within which the cluster's density is 500 times the critical density at the cluster's redshift ($M_{500}$; \citealt{PlanckCollaboration2016}), 
    central entropy ($K_0$; \citealt{Giacintucci2017}), X-ray morphological parameter \citep[$c,w$;][]{Campitiello2022}.}
    \label{tab:properties}
    \begin{center}
    \setlength{\extrarowheight}{3pt}
    \begin{tabular}{ L R }
    \toprule\toprule
    \text{z}                          & 0.1427         \\  
    \text{R.A. (h, m, s)}             & 11\, 55\, 18.9 \\ 
    \text{Dec }(^\circ,','')          & +23\, 24\, 31  \\
    M_{500}\,(10^{14} \rm M_{\odot})  & 5.95^{+0.24}_{-0.25} \\
    K_0\,(\rm keV \, cm^2)            & 64\pm8 \\ 
    c                                 & 0.44^{+0.04}_{-0.04} \\
    w                                 & 0.04^{0.02}_{0.01}\,\cdot 10^{-1} \\
    \bottomrule
    \end{tabular}
    \end{center}   
\end{table}

\begin{figure*}
\centering
    \begin{subfigure}{0.49\textwidth}
      \centering
      \includegraphics[width=\textwidth]{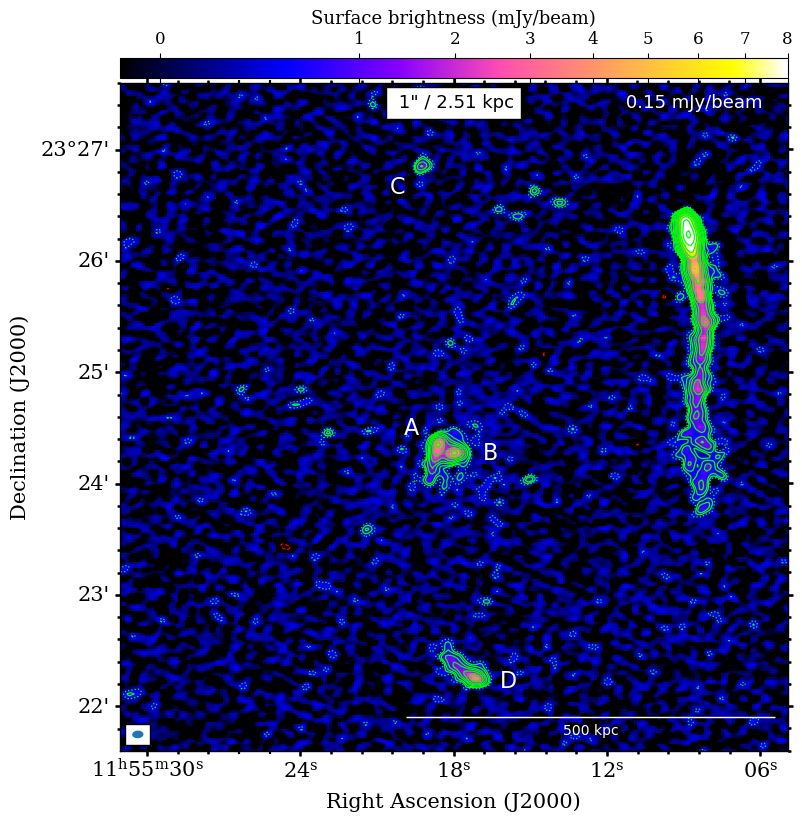}
      \label{fig:HR_radio}
    \end{subfigure}%
    \begin{subfigure}{0.49\textwidth}
      \centering
      \includegraphics[width=\textwidth]{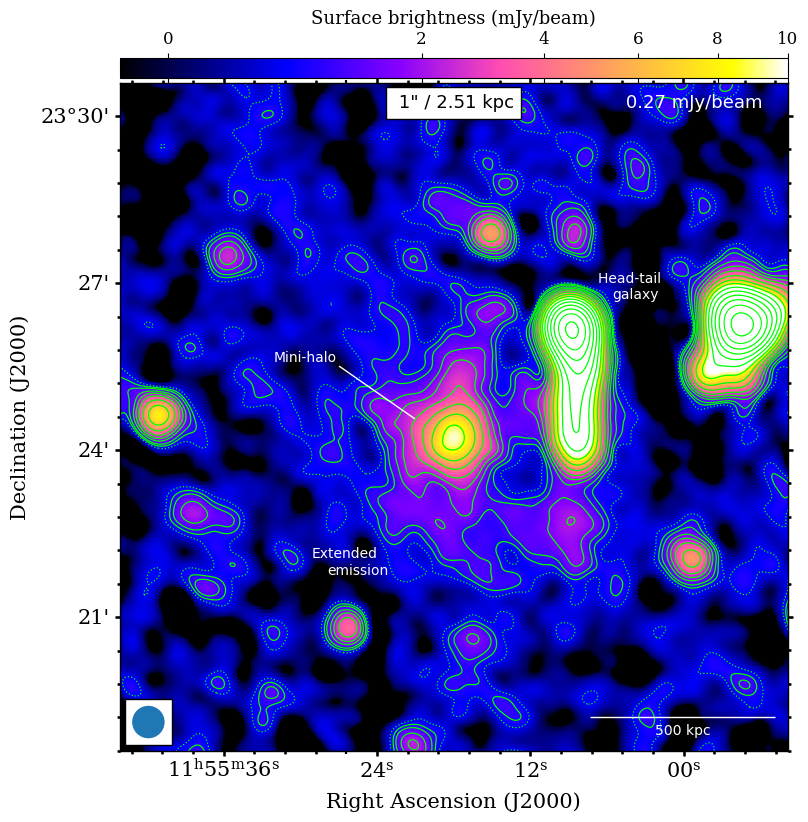}
      \label{fig:LR_radio}
    \end{subfigure}%
  \caption{High and low resolution radio images of A1413 at 144 MHz.
  Left panel: High resolution 144~MHz image with $\mathrm{4''\times6''}$ beam (shown in the bottom left corner) and noise $\mathrm{\sigma_{\rm rms}=0.15 \,m Jy/beam}$. 
  Sources that have been subtracted are labeled with letters: two central AGN A, B; one point-like north source C and one extended south source D. 
  Right panel: Low resolution source-subtracted 144MHz image with  $\mathrm{35''\times35''}$ beam (shown in the bottom left corner) and noise $\mathrm{\sigma_{\rm rms}=0.27 \,m Jy/beam}$. 
  For both images, the contour levels start at $2\sigma_{\rm rms}$ and are spaced with a factor of $\mathrm{\sqrt{2}}$. The $-3\sigma_{\rm rms}$ contours are red coloured, while $2\sigma_{\rm rms}$ contours are dotted green coloured.
    }
     \label{fig:radio_lowfreq}
\end{figure*}

\section{Observations and data reduction}
\label{sec:datared}

In order to investigate the radio emission and dynamical state of A1413 we used LOFAR data at 144 MHz together with newly analysed archival data from XMM-Newton telescope, the latter being also part of the CHEX-MATE program \citep{CHEX-MATECollaboration2021}.
Moreover, we compare the LOFAR observation of the source with VLA image of this source presented in \cite{Govoni2009}.

In the following sections, we describe these observations and the data reduction procedures. A summary of the observational details is reported in \cref{tab:Observation}.

\begin{table*}[ht]
\captionsetup{margin=2em}
\caption{Observational overview: LOFAR HBA and VLA observations (top); X-ray observation (bottom).}
\label{tab:Observation}
\begin{center}
\setlength{\extrarowheight}{3pt}
\begin{tabular}{ c c c c c c c  }
 \toprule\toprule
 
 Telescope & Central frequency & Configuration & On source time & Observation ID & Observation Date & Reference \\ \midrule
 
 LOFAR & 144 MHz   & HBA\_DUAL & 8+8 hrs & 591537,602890 & 04.05.2017,09.08.2017  & (1) \\  
 VLA   &  1.4 GHz  & C array  &  8 hrs  &     & 23.10.2006   & (2)  \\ \bottomrule\bottomrule
 Telescope & Energy range & Filter & On source time & Observation ID & Observation Date  & Reference \\ \midrule
 \XMM & [0.5-14] keV & THIN1 & 82.5 ks & 0502690201 & 11.12.2007, rev. \# 1466 & (1),(3)\\
 \bottomrule
\end{tabular}
\end{center}
\vspace{-0.3cm}
\captionsetup{justification=centering}
\caption*{References: (1) this work; (2) \cite{Govoni2009}; (3) \cite{Pratt2002}.}

\end{table*}
%
\subsection{\XMM}\label{subsec:datared_xray}

A1413 was observed with \XMM four times between 2007 and 2008: all four observations are longer than 60 ks each. For this work, we retrieved the longest and less flared one, OBS ID 0502690201, from the XMM-Newton data archive (see \cref{tab:Observation}). 
Data consists of 82.5 ks archival observation time during XMM-Newton revolution 1466 (2007 December 11), obtained with the THIN1 filters and are processed with XMM-SAS\footnote{\url{https://www.cosmos.esa.int/web/xmm-newton/sas}} v19.0.0 and the XMM-ESAS\footnote{\url{https://heasarc.gsfc.nasa.gov/docs/xmm/xmmhp_xmmesas.html}} software package \citep{Snowden2008}, through a dedicated pipeline that follows the procedure described in \cite{Lovisari2019} and which is briefly described in this section. The pipeline processes the X-ray data from the raw data, the Observation Data File (ODF), 
passing through data calibration and filtering, imaging, and spectral fitting.

The raw data are initially processed using the  \texttt{\footnotesize cifbuild} and \texttt{\footnotesize odfingest} task. The former provides an up-to-date list of the relevant calibration files, while the latter produces extended ODF summary file needed for subsequent data processing. 
Then we run \texttt{\footnotesize emchain} and \texttt{\footnotesize epchain} tasks to convert the raw data into calibrated photon event files for the MOS and pn camera, respectively. 
We applied standard filtering procedures to discriminate events based on their pattern: only single, double, triple and quadruple events for MOS (i.e. PATTERN $\leq12$) and single and double for pn (i.e. PATTERN $\leq4$).
We also performed bright pixels and hot columns removal (i.e. FLAG==0) and removed the contamination by pn Out-of-Time events. 
Observation periods affected by high background due to soft protons (SP; see \cref{subsubsec:datared_background}) were removed using a two-stage filtering process (see \citealt{Lovisari2011} for details), summarised in the following. 
First, the light curve with 100 s bins is inspected in [10-12] keV band for MOS and [12-14] keV for pn, since particle background dominates at high energy. The histogram of the light curve is then fitted with a Poisson distribution and time intervals with a higher count rate, i.e. a count rate that deviates by more than 2$\sigma$ from the mean, are excluded (see \citealt{Pratt2002}, Appendix A for a precise description). To obtain the final Good Time Intervals (GTI), a second-stage filtering is performed, this time using the full [0.3-10] keV band in 10 s bins, as a safety check for possible flares with soft spectra \citep[e.g.][]{Nevalainen2005}.

Filtering the light curve does not ensure to delete completely the SP contributions: continuous and constant flares (i.e. flares that last for long time intervals), without evident spikes, can still remain. We therefore estimate the amount of residual SP flare contamination by using the suited code of \cite{DeLucaMolendi2004}. 
By comparing area-corrected count rates in the in-FOV and out-of-FOV regions of the detector, we find that the SP contamination is low\footnote{\url{https://www.cosmos.esa.int/web/xmm-newton/epic-scripts}} with a ratio of $1.12\pm0.02$ for all the cameras.
Nevertheless, we included SP contamination as an extra model component, when dealing with spectral fitting (\cref{subsubsec:datared_spectralanalysis}).
Anomalous behaviour of CCD 7 for MOS2 was detected via \texttt{\footnotesize mos-filter} task \citep{Kuntz2008} and excluded from this analysis, together with CCD 6 of MOS1, that was damaged by a meteorite in the early phase of the emission and thus, not available.

Point-like sources were detected with the \texttt{edetect-chain} task and visually inspected to discriminate between real point sources and extended cluster substructures before removing them from the event files.
Even if identified as point-source, we did not remove the BCG region because the cluster emission is so bright that the contribution of the AGN is negligible and can be modeled as a power law during the spectral fitting.
\subsubsection{Background treatment}\label{subsubsec:datared_background}
\XMM EPIC background is made of several different components, each one with temporal, spectral, and spatial variations. It can be generally divided between particle and photon background. 

%
The main component of the photon background is the cosmic X-ray background (CXB), primarily consisting of unresolved cosmological sources, the Local Bubble and the Galactic Halo.
To model the CXB we used ROSAT All-Sky Survey\footnote{Since in ROSAT observations, particle background is negligible, these can be considered as pure sky data. RASS diffuse background maps can be downloaded at the HEASARC webpage \url{https://heasarc.gsfc.nasa.gov/cgi-bin/Tools/xraybg/xraybg.pl}.} (RASS) diffuse background maps in a region beyond the cluster virial radius (between 1 and 2 degrees) and fitted it simultaneously with \XMM data, as described in \cite{Lovisari2019}.
We described the unresolved emission from AGN via an absorbed power-law model with its slope set to 1.41 \citep{DeLucaMolendi2004}; an absorbed $\sim0.2$ keV thermal  component  representing  the  Galactic  Halo (GH)  emission, and an unabsorbed  $\sim0.1$ keV component representing the Local Hot  Bubble  (LHB).
During the fitting procedure, the temperature values of the LHB and GH thermal components are free to vary, while the slope of the AGNs' power law is fixed.

%
The particle background includes signal generated by interactions of particles with the detectors and depends on the energy of the hitting particles.
It can be divided between the quiescent particle background (QPB, with $\rm E \gtrsim few\,MeV$) and soft protons (SP, with $\rm E < 100\,keV$). 
QPB consists of a continuum component and fluorescent lines produced by the interaction of penetrating high energy particle with the detectors. We used Filter Wheel Closed\footnote{FWC exposures are taken with closed filters that avoid X-ray photons to hit the detector and thus, are dominated by the instrumental background.} (FWC) observations to estimate the intensity of QPB components. 
The model used for FWC observations consists in a broken power-law with additional fluorescence lines (e.g. Fe,  Al, Si, Au). 
Since FWC observations are taken in different time period from our observation and have a different exposure, they need to be normalised, comparing them with the out-of-field-of-view data (unexposed corners of the CCD chips are a measure of the particle background level in each observation.).
SP background is produced by relatively low energy protons accelerated in the Earth magnetosphere, passing through the telescope optics and depositing their energy directly on the CCDs.
As described in \cref{subsec:datared_xray}, SP components are usually highly variable in time (ranging from $\rm \sim100\,s$ to hours), and the enhancement in the count rate (referred to as "flares") can be more than three orders of magnitude. This flare component is usually filtered via light-curve screening procedures, summarised in \cref{subsec:datared_xray}.
Nonetheless, a more constant SP components can remain and contaminate the observations. To take into account this residual SP contamination, we added a power-law component, folded only with the Redistribution Matrix File, to the background modelling. We left both the slope and normalisation free to vary for each detector and in the region of interest. This approximately accounts for the proton vignetting (which differs from the photon one, and for which there is no model available).
%
%
%
\subsubsection{Spectral analysis }
\label{subsubsec:datared_spectralanalysis}
The spectral analysis is performed with the \texttt{XSPEC} \citep{Arnaud1996} package v12.11.1, in  the  [0.5-12] keV and [0.5-14] keV energy range for MOS and pn, respectively. 
All the regions used for the spectral profiles were centred on the peak of the X-ray emission of coordinates R.A.(J2000) $11^{h} 55^{m} 17^{s}.85$, DEC. (J2000) $+23^{\circ} 24^{m} 18^{s}.85$, and extend until $\rm R_{500}\sim 1224\,kpc$ \citep{PlankCollab2014}. 
The size of the annuli have been determined by requiring a minimum width of $30''$ and a fixed S/N=50. The first requirement ensures that most of the flux \citep[i.e. $>80$\%,][]{Zhang2009} comes from the selected region (due to the XMM–Newton PSF, some photons scatter from one annulus to another); the second that the $\mathrm{1\sigma}$ uncertainty in the spectrally resolved temperature (and consequently in the fitted temperature profiles) is <10\% (see Appendix B of \citealt{Lovisari2019} for more details of how uncertainties vary as function of S/N and cluster temperature). 
The cluster emission is modelled with an \texttt{apec} single-temperature thermal plasma model with an absorption fixed at the total \citep[neutral and molecular; see][]{Willingale2013} $\rm N_H\sim 1.97\cdot10^{20}\,cm^{-2}$ value estimated using the SWIFT online tool\footnote{\url{https://www.swift.ac.uk/analysis/nhtot/index.php}}. For the \texttt{\footnotesize apec} model \citep{Smith2001} the parameters are: plasma temperature [keV], metallicity, redshift (that is kept fixed at z=0.1427), and the normalisation $\frac{10^{-14}}{4\pi[D_A(1+z)]^{2}}\int n_e n_H dV$, with $D_A$ the angular distance of the source, $n_e$,$n_H$ the electron and proton densities in units of $\rm cm^{-3}$. 
In each annulus, the MOS and pn spectra were fitted simultaneously and their normalisations left free to vary, in order to account for the calibration offsets between the different detectors \citep[e.g.,][]{Madsen2017}.
The background is modelled together with the cluster emission by performing a joint fit with the above explained RASS spectrum and FWC observations components, and cluster emission extracted from each region. 

\subsection{LOFAR}\label{subsec:datared_radio}

The data used in this work are part of the LOFAR Two-metre Sky Survey \citep[LoTSS;][]{Shimwell2017, Shimwell2019,Shimwell2022}, a sensitive ($\rm rms\sim100\,\mu Jy/beam$), high-resolution (FWHM of the synthesised beam of $\mathrm{\sim6"}$), low-frequency (120-168 MHz) radio survey of the northern sky. 
LoTSS observations are conducted with LOFAR (LOw Frequency Array: \cite{vanHaarlem2013}) High Band Antennas (HBA), in the \texttt{DUAL\_INNER} mode configuration, each pointing consists of 8 hrs observation book-ended by 10-min scans of the calibrator ($\rm 3c395$ for the observations used in this work), at the central frequency of 144 MHz and 48 MHz total bandwidth.
A1413 is covered by two pointings, P177$+$22, and P180$+$22. 

Data have been calibrated following the LoTSS-DR2 pipeline  (see \citealt{Shimwell2022, Tasse2021} for details), which 
works on the direction-independent calibration 
products (i.e. 24 measurement sets for each 
pointing, that cover the 120-168 MHz band) and 
returns fully direction-dependent calibrated data. %
After calibration, the so called "extraction and self-calibration" procedure \citep{vanWeeren2021extraction} is performed: a region around the target of interest is selected and all the sources outside that region are subtracted using the direction-dependent gains obtained by the pipeline. The station beam is corrected for directly in the visibility space, and the phase center of the observation  is shifted toward the direction of the target. This way, we are left with 2 data sets centered on the target that do not need any further beam correction. 
These 2 data sets are self-calibrated again to obtained more accurate calibration solutions in the region of interest.
Self-calibration and imaging are done with DPPP and WSClean 2.10.0 \citep{Offringa2014}.

We have re-imaged the self-calibrated data sets at different resolutions to classify the kind of radio emission and its properties. 
We always applied a Briggs \citep{Briggs1995PhDT} weighting scheme with a negative robust parameter, as it allows to down-weigh the secondary lobes, and inner $uv$-cut fixed at $\mathrm{b_{min}>80\lambda}$ to avoid interference on short baseline, i.e. to filter out emission on angular scale larger than $43\arcmin$. This is a standard procedure used in LoTSS analysis \citep[e.g.,][]{Botteon2022DR2}.
We performed images at different resolutions: high-resolution (HR) images to subtract point-sources which overlaps to the diffuse emission and low-resolution (LR) images to gain sensitivity to the diffuse emission (see \cref{tab:radio_images_details} for different imaging parameters used in this work). 

%
%
On the HR image (\cref{fig:radio_lowfreq}, left) we identified different sources, named with letters (A, B, C, D), within a radius of 500 kpc ($\rm \sim0.5 R_{500}$) from the cluster centre. The emission from these sources contaminates the diffuse emission that we want to study, and thus it has been subtracted. 
The tailed galaxy on the right, with optical counter part at $z = 0.144$, is instead quite extended and difficult to remove and disentangle from the diffuse radio emission.
Therefore, we decided to not subtract it. 
The source subtraction is performed directly from the visibilities through a multi-step procedure.
First, we produced images at high resolution ($5\arcsec$ FWHM) to identify these compact sources (A, B, C, D in \cref{fig:radio_lowfreq}, left). We selected the corresponding clean components in the model image and, through an anti-Fourier transform, converted them into visibilities. Finally, a new data set is created as subtraction of the selected model components $uv-$data from the original data.
Given the slight extension of source A and D, we performed an additional subtraction step, imaging the subtracted data sets at an intermediate $\sim10\arcsec$ resolution. 
This allowed to definitively remove all the sources, as they do not leave residual when inspecting the final $uv-$subtracted data set.
%
Eventually, the final data set is imaged maintaining a negative Briggs weighting with $\mathrm{R=-0.25}$ and tapering down the baselines longer than $\mathrm{10\,k\lambdaup}$ (i.e. using a Gaussian taper with a FWHM of 20\arcsec) to gain sensitivity to the diffuse emission (see \cref{fig:radio_lowfreq}, right).

To ensure a fair comparison with other LOFAR works on the same cluster, we point out that we used different observations respect to the previous LOFAR study of \cite{Savini2019}. In particular, these authors used a single 8-hours pointed observation, while we combined two pointings for a total of 16-hours observation. Moreover, we used improved calibration techniques that allowed to produce images with higher sensitivity. For example, our $15\arcsec$-resolution image has $\rm \sigma_{rms}$ improved more than a factor of two respect to the one of \cite{Savini2019}, with $\rm \sigma_{rms}=450~\mu Jy/beam$ at $\sim 20\arcsec$-resolution.
On the other hand, \cite{Riseley2023} used the same set of observations presented here and achieved comparable sensitivity. However, these authors undertaken an independent post processing.

Throughout the paper, the error $\Delta_S$ associated with a flux density measurement, $S$, is estimated as
\begin{equation}
    \Delta_S = \sqrt{  (\sigma_{c}\cdot S)^2 + N_{\rm beam}\cdot \sigma_{\rm rms}^2} ,
\end{equation}
where $\rm N_{beam}=N_{pixel}/A_{beam}$ is the number of independent beams in the source area, and $\rm \sigma_{c}$ indicates the systematic calibration uncertainty of the flux density, with a typical value of 10\% for LOFAR HBA observations \citep{Shimwell2022}. 
Moreover, we point out that different statistical significance is used, depending on the kind analysis. In the assessment of the radio diffuse emission's proprieties, such as size, flux density, and power, we employ the standard $3 \sigma_{\rm rms}$ detection limit (see \cref{subsec:radio_morphology}, \cref{subsec:MH_integrated_spectral_index}). This also allows fair comparison with other works on the same cluster (e.g. \cite{Govoni2001}, \cite{Riseley2023} cited in the paper). 
However, as explained in \cref{subsec:spatialcorrelation}, when performing surface brightness measurements, a $2 \sigma_{\rm rms}$ is applied to avoid biased results.

\begin{table*}
    \centering
    \renewcommand\arraystretch{1.5}
    \captionsetup{justification=centering}
    \caption{Radio images used in this work.}
    \vspace{0.3cm}
    \label{tab:radio_images_details}

    \begin{tabular}{ c c c c c c l  }
    \hline
    \hline
    
    Telescope & Freq  & Briggs weighting & Taper FWHM  & Resolution                & $\rm \sigma_{rms}$ & Reference \\
              & [MHz] &                   & $[\arcsec]$ & [\arcsec$\times$\arcsec]  & [mJy/beam]         &           \\ \hline
    HBA       & 144   & $-1$           & -                 & $4\times6 $        & 0.15   &  \cref{fig:radio_lowfreq} (left) \\  
    HBA       & 144   & $-0.25$        & $20$              & $35\times35$       & 0.27   &  \cref{fig:radio_lowfreq} (right)  \\
    HBA       & 144   & $-0.25$        & $10$              & $15\times15$       & 0.18   &  \cref{fig:HBA_15asec_and_annuli} (left) \\
    \hline
    \end{tabular}
\end{table*}

\section{Results}\label{sec:results}
As presented in \cref{subsec:A1413}, A1413 is a peculiar galaxy cluster, showing intermediate characteristics between relaxed and merging clusters. Past radio observations have revealed a mini-halo at its centre \citep{Govoni2009, Savini2019}.
In this section, we present the main results of our X-ray and radio analysis. We used shape of the surface brightness (SB), temperature, and  metallicity profiles, as well as the central cooling time to determine the cluster's dynamical state. Moreover, we analysed the radio surface brightness in 1D and 2D to classify the morphology of the radio source and define its properties. 

\begin{figure*}[ht]
\centering
%
 \captionsetup{justification=centering}
 \begin{subfigure}{0.45\textwidth}
     
     \centering
     \includegraphics[width=\textwidth, height=7cm]{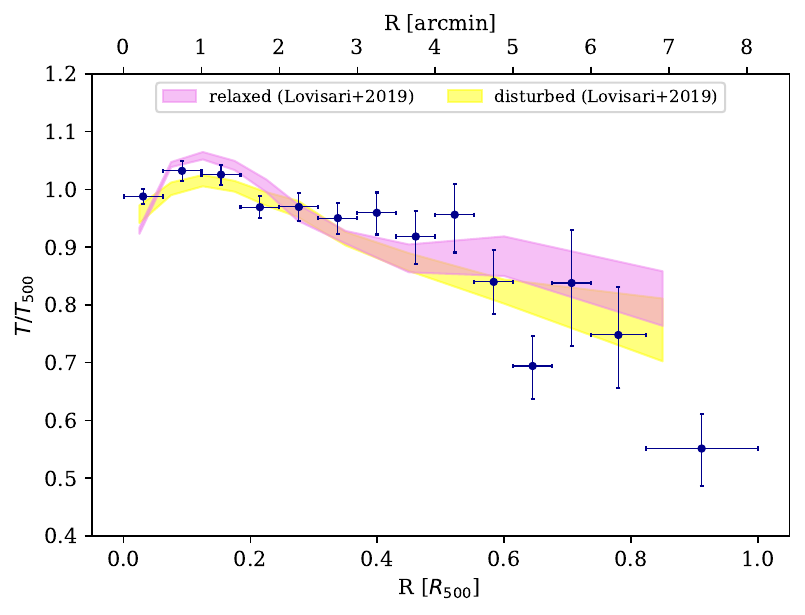}
 \end{subfigure}
 \begin{subfigure}{0.45\textwidth}
     \centering
     \includegraphics[width=\textwidth, height=7cm]{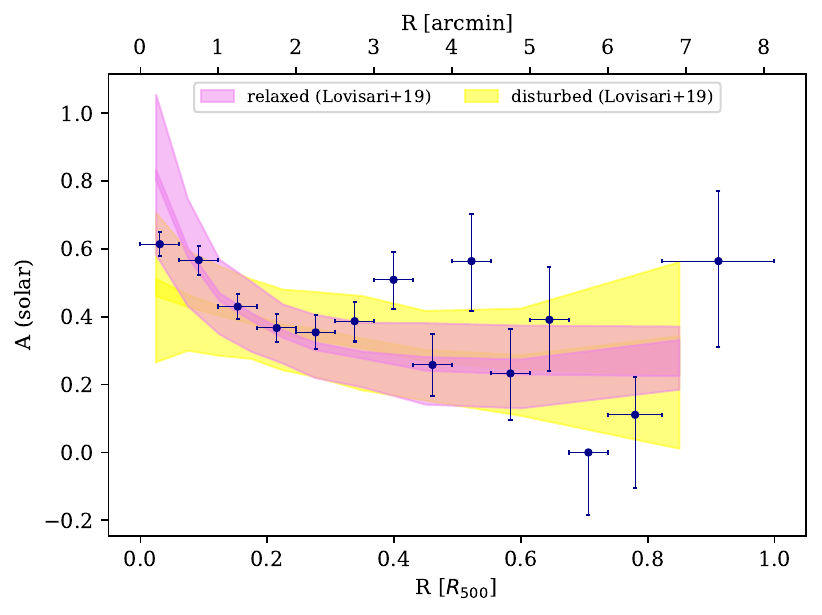}
 \end{subfigure}
\captionsetup{justification=justified}  
\caption{
1D thermodynamical profiles obtained from spectral fitting, up to $R_{\rm 500}$.  
Left panel: Projected temperature profile. Right panel: Metals abundance profile. Both profiles are compared with different averaged values of relaxed systems sample (pink) and disturbed ones (yellow) from \cite{Lovisari2019}. 
}
 \label{fig:specfit_results} 
\end{figure*}

\subsection{X-ray azimuthal spectral profiles}

The projected temperature profile is shown in  \cref{fig:specfit_results} (left). 
It is well known that for relaxed cool-core galaxy clusters, the temperature profile shows a clear drop in the central region \citep[e.g.,][]{Vikhlinin2005, Pratt2007}. However, for A1413, the temperature does not show the drop in the central region clearly: just the innermost bin ($0.5\arcmin\sim75\,\rm kpc$ width) shows a drop from $7.43\pm0.12$ to $7.11\pm0.01$ keV. On the other hand, at larger radii the temperature profile decreases as expected for typical, relaxed cluster. This is consistent with the analysis of \cite{Pratt2002}, even if we point out that it could be biased by the low resolution of XMM \citep[as pointed out in][]{Vikhlinin2005}.\\
%
%
%
We also obtained the metal abundance of the \texttt{\footnotesize apec} model, which is a mean abundance gathering the possible emission from different elements (e.g., C, N, O, Ne, Mg, Al, Si, S, Ar, Ca, Fe, Ni), where Fe generally dominates. The profile is displayed in Fig.~\ref{fig:specfit_results} (right) and does not present clear peak in the cluster core, typical of relaxed objects. 
We compared the observed profiles to average temperature and metallicity profiles for relaxed (pink) and disturbed (yellow) galaxy groups and clusters studied by \cite{Lovisari2019}, as a reference. 
Fig.~\ref{fig:specfit_results} shows that both properties do not clearly follow either the mean profile of relaxed nor of disturbed clusters.
In the core, where actually relaxed and disturbed cluster profiles differ, A1413 sits in between the two distributions. This supports the scenario that A1413 is an intermediate phase system.
\subsection{X-ray $-$ optical offset}
For dynamically relaxed systems, we expect the BCG to be at the centre of the gravitational potential well \citep{vandenBosch2005}. 
Several X-ray studies have proven the connection/correlation between the X-ray peak$-$BCG offset and the disturbed dynamical state \citep[e.g.,][]{Hudson2010,MannEbeling2012, Rossetti2016, Pasini2021}, making this indicator a robust and reliable diagnostic for assessing the dynamical condition of clusters.
We find a small offset of about $7.2$ kpc between the X-ray peak with the optical BCG.
The coordinates of the central galaxy were available R.A. (J2000) $11^{h} 55^{m} 17^{s}.87$, DEC. (J2000) $+23^{\circ} 24^{m} 16^{s}.02$ \citep{Rawle2012} and correspond to the brightest pixel of the optical image, retrieved from HST (F606W filter) archive. 
Despite this offset, recent statistical studies by \cite{Rossetti2016}, show that an offset less than $0.02\,R_{500}$ (as in this case $\rm 7.2 \,kpc=0.006\,R_{500}$) is still consistent with a relaxed system. Moreover we point out, this offset corresponds to angular size of $\sim 3\arcsec$, which is slightly bigger than the X-ray image's pixel size ($\sim2\arcsec$), used to determine the X-ray peak.\\
%
%
%
%
%
\subsection{X-ray surface brightness profile}\label{subsec:SB}
\begin{figure}
  \centering
  \includegraphics[width=0.49\textwidth]{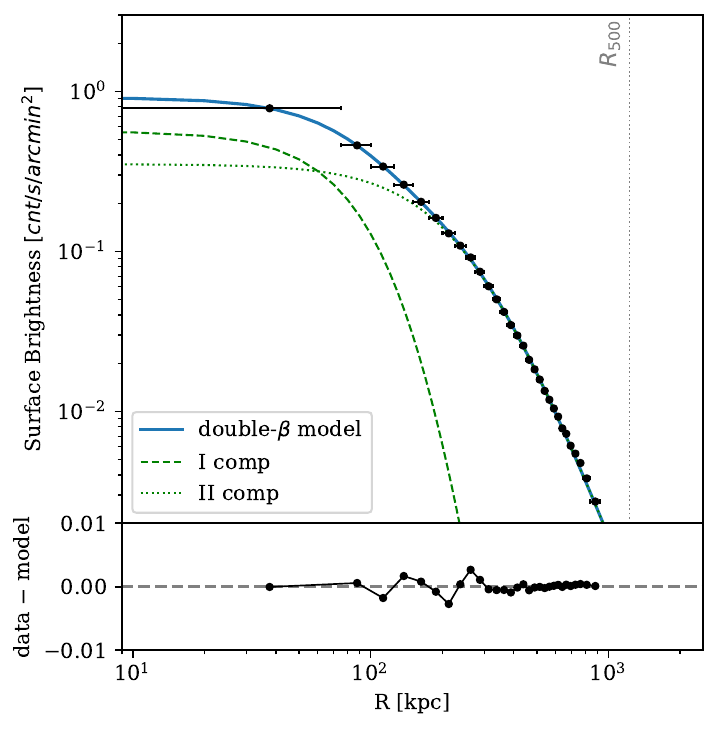}
  \caption{X-ray surface brightness profile, together with the best fit double-$\beta$ model of \cref{eq:double_beta}. The best fit parameters are $\Sigma_{01} = 0.56\pm 0.05\rm\, cst/s/arcmin^{2}$, $r_{c1} =189.62\pm 0.41\rm \,kpc$, $\beta_1 =2.17\pm 0.91$ and $\Sigma_{02} = 0.35\pm 0.04\rm\, cst/s/arcmin^{2}$, $r_{c2} = 257.04\pm 0.25\rm \,kpc$, $\beta_2 = 0.80\pm 0.04$ with $\chi^2/d.o.f \simeq1.64$. The vertical line indicates $R_{500}$. The surface brightness values have errors of the order of 1$-$2\% , thus error bars on y-axis are not visible in this plot.}
  \label{fig:SB_Xray}
\end{figure}

We computed the radial surface brightness profiles (SB) in the [0.7–2] keV energy band (\cref{fig:SB_Xray}). This is the energy range where, for temperatures as those observed in A1413, the SB has a very small dependence on the $kT$ and it also provides an optimal ratio of the source and background flux in \XMM data. The annuli for the SB profile are again centred on the X-ray peak and have been determined by requiring a fixed $\rm S/N=10$ and minimum width of $2''$.

The profile is fitted with a double $\beta$-model \citep{LaRoque2006} in the form

\begin{equation}
    \Sigma = \Sigma_{01} \Bigg[ 1+ \Big(\frac{r}{r_{c1}} 
    \Big)^{2} \Bigg]^{-3 \beta_{1}+1/2}   +   \Sigma_{02} \Bigg[ 1+ \Big(\frac{r}{r_{c2}} 
    \Big)^{2} \Bigg]^{-3 \beta_{2}+1/2} ,
    \label{eq:double_beta}
\end{equation}
where $r_{c1}$,$r_{c2}$ are the core radii, $\Sigma_{01}$, $\Sigma_{02}$ are the central surface brightnesses of the two components and $\beta$ represents the ratio of specific kinetic energies of galaxies and gas \citep[e.g.,][]{Gitti2012}. 
This is the extension  of the isothermal $\beta$-model \citep{CavaliereFusco1976}, commonly used to describe the surface brightness profile of ICM. In fact, when fitting the profile over the radial range with a single $\beta$-model, we found the presence of a central excess. This biases the best fit results toward a small core radius and overestimates
the true profiles in the region outside the excess ($r\gg r_c$), producing non-random strong residuals.
%
Thus, a second component was needed and it is typically required to reproduce the sharp X-ray peak in the core of relaxed clusters \citep[e.g.,][]{Mohr1999, Pratt2002, Pasini2019}.
%
%
%
\subsection{Cooling time}\label{subsec:cooling_time}
\begin{figure}
  \centering
  \includegraphics[width=0.49\textwidth]{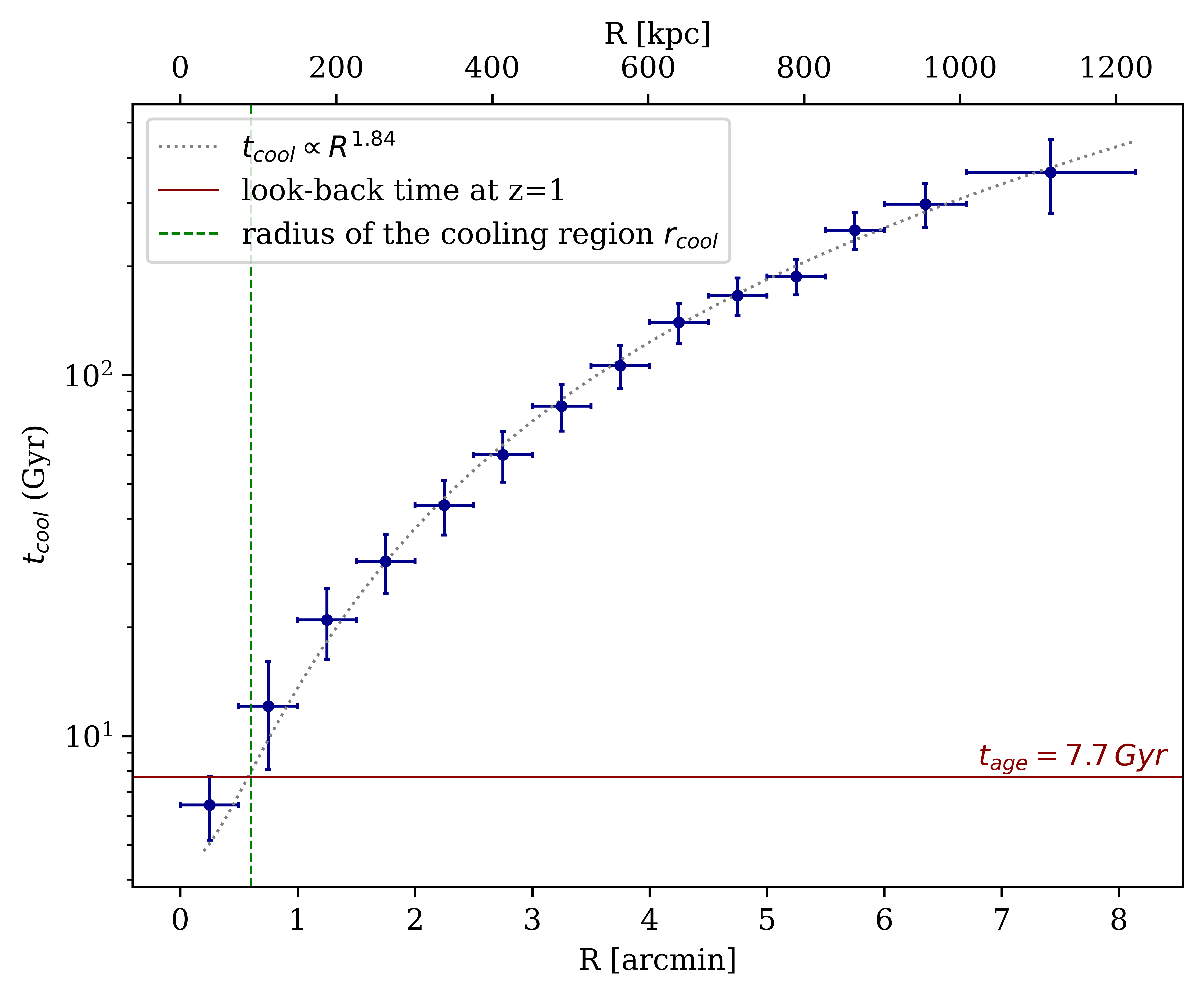}
     \caption{Cooling time profile for A1413. The red solid line represents the look-back time at z=1,  $\rm t_{age}=7.7 \, Gyr$.  The power-law fit with $\rm t_{cool}=9.27 R^{1.84}$ allows to define the cooling radius (green vertical line), namely the region within which $\rm t_{cool}<t_{age}$.}
     \label{fig:cooling_profile}
\end{figure}

From the temperature and density profiles, the cooling time of each region can be estimated as
\begin{equation}
    t_{\rm cool} = \frac{H}{\Lambda(T)n_e n_p} = \frac{\gamma}{\gamma -1} \frac{kT(r)}{\mu X n_e(r) \Lambda(T)} ,
    \label{eq:cooling_time}
\end{equation}
where $\gamma=5/3$ is the adiabatic index, $H$ is the enthalpy, $\mu\sim0.6$ is the molecular weight for a fully ionised plasma and $\Lambda(T)$ is the cooling function \citep{StutherlandDopita1993}. 
Given the cooling profile, we can to define the cooling radius, i.e. the radius at which the cooling time is shorter than the age of the system. The latter can be estimated assuming the cluster's age to be equal to the look-back time at $z=1$, that is, $t_{\rm age}\simeq 7.7$ Gyrs \citep[e.g.,][]{Birzan2004, Gitti2012}, as at this time many clusters appear to be relaxed (\cref{fig:cooling_profile}, red line). 
To obtain $r_{\rm cool}$ we qualitatively fitted the cooling time profile with a power-law relation and selected $r_{\rm cool}$ as the intersection between the best-fitting power law and $t_{\rm age}$. Thus, the cooling region for A1413 is within $r_{\rm cool}= \rm 0.6 \,arcmin = 90\, kpc $ (\cref{fig:cooling_profile}, green vertical line). 

\cite{Hudson2010} argued that the central cooling time $t_{\rm cool}$ is the best parameter, for low-redshift clusters, to identify cool-core and non-cool-core clusters.
They divided clusters into three types: $t_{\rm cool}<1$ Gyr for strong cool-core clusters; $t_{\rm cool}\sim\,1-7.7$ Gyr for weak cool-core clusters; $t_{\rm cool}>7.7$ Gyr for non-cool-core clusters. Following the definition of a cool-core based on the cooling time given by \cite{Hudson2010}, A1413 appears to be a weak cool-core cluster with a central cooling time of $t_{\rm cool}= 6.4\pm1.3$ Gyr within $r<0.5\arcmin\sim75\,\rm kpc$.\\

%
\subsection{Radio morphology}
\label{subsec:radio_morphology}
Previous studies of A1413 at 1.4 GHz \citep{Govoni2009} and 144 MHz \citep{Savini2019} frequencies confirmed the cluster hosting a central mini-halo. 
Thanks to the improved calibration strategy and our deeper LOFAR HBA observations, we detect low surface brightness emission which extends further out from the cooling region, only visible in the LoTSS images. 
It is elongated in the North-South direction, as it happens for the X-ray, and has a maximum linear size of $\rm D\sim800\,kpc$ ($3\sigma_{\rm rms}$-contour), on scales typical of giant radio halos (see source-subtracted $35\arcsec$-resolution image of \cref{fig:radio_lowfreq}, right). 
West of the cluster, a head-tail radio galaxy with optical counter part at $z = 0.144$ (i.e., consistent with the cluster's redshift) is present, which appear to be connected with the outer radio emission, though we can not exclude projection effects. 
Given the new discovery, we analysed the radio emission of the cluster to understand whether we detected a giant radio halo or as the superposition of two different sources (mini-halo plus halo emission), as recently found in few other clusters (RX J1720.1+2638 of \cite{Biava2021}, Abell 2142 of \cite{Bruno2023_A2142}).
%
%
\subsection{Radio surface brightness profiles}\label{subsubsec:radio_SB}
%
%
The exponential profile has been customarily used to fit the SB profile of giant and mini-halos \citep[e.g.][]{Murgia2009,Botteon2022DR2,Bonafede2022,Cuciti2022Nat}.
This is a simple model with only two free parameters, and it provides a reasonable description of many radio halos.
We studied the radio surface brightness making use of two approaches: we first performed a 1D analysis following the fitting procedure of \cite{Murgia2009}, and then the more recent 2D method developed by \cite{Boxelaar2021}.

However, we point out that this is not a physical model, and \cite{Botteon2022DR2, Botteon2023} showed that the exponential is not always a good model for representing halos. In particular, when halos present substructures in the radio emission (such as edges and filaments), the exponential smooth profile leaves strong residuals.
\\

\begin{figure*}
\centering
    \begin{subfigure}{0.45\textwidth}
      \centering
      \includegraphics[width=\textwidth]{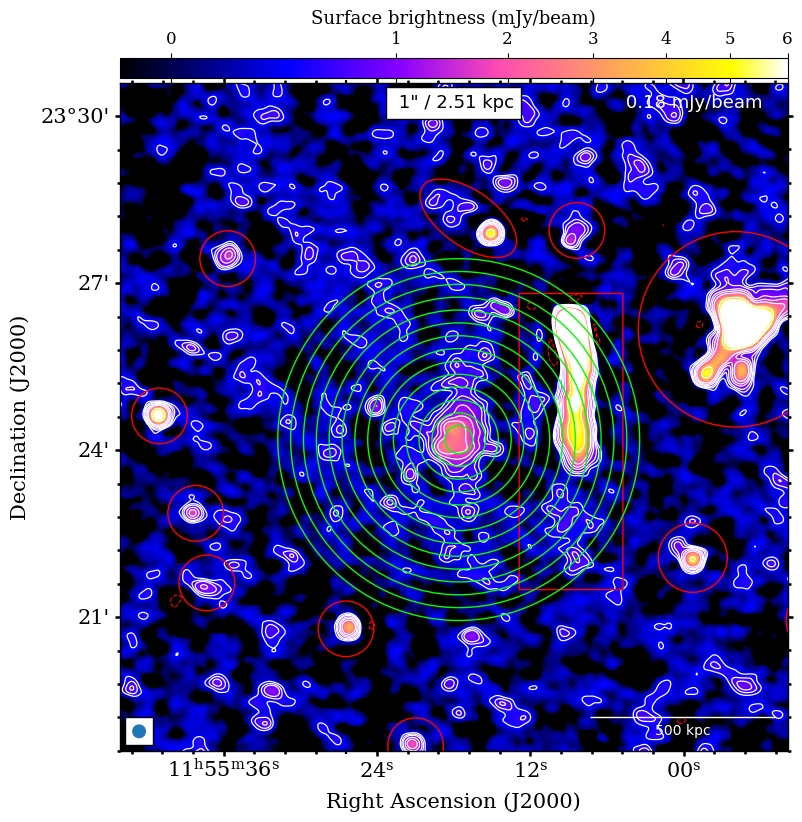}
    \end{subfigure}%
    \begin{subfigure}{0.45\textwidth}
      \centering
      \includegraphics[width=\textwidth]{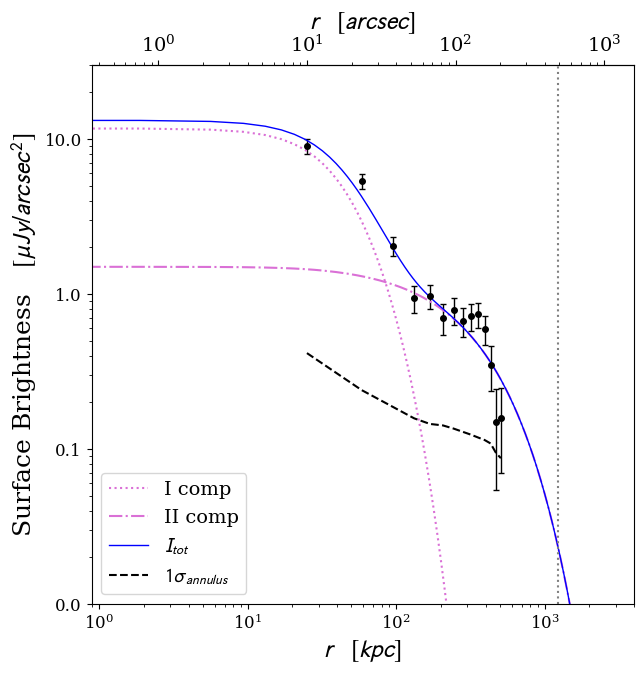}
    \end{subfigure}
\caption{
1D radio surface-brightness profile.
Left panel: Medium resolution ($15\arcsec\times15\arcsec$) source-subtracted image at 144 MHz. Masked regions are shown in red, together with the green annuli where the surface brightness has been extracted.
The contour levels start at $2\sigma_{\rm rms}$, where  $\mathrm{\sigma_{\rm rms}=0.18 \,m Jy/beam}$, and are spaced with a factor of $\mathrm{\sqrt{2}}$. The $-3\sigma_{\rm rms}$ contours are red coloured.
Right panel: Radio surface brightness profile. The solid line represents the resulting double exponential model, while the single components are plotted as reference. The central mini-halo component has $r_{\rm e,1} = 28\pm5$ kpc, while the second broader component $r_{\rm e,2} = 290\pm60$ kpc. The vertical line indicates $R_{500}$ and the dashed black line the $1\sigma_{\rm annulus}$ detection limit for each annulus.}
\label{fig:HBA_15asec_and_annuli}
\end{figure*}

Following \cite{Murgia2009}, we calculated the radio brightness profile, averaged in concentric annuli of $\mathrm{15\arcsec}$ width, i.e. the FWHM of image beam, centred on the peak of the radio halo (\cref{fig:HBA_15asec_and_annuli}). In fact, to increase the number of points for the fit, we used the source-subtracted $\mathrm{15\arcsec}$ resolution image, which is the most sensitive to the central mini-halo emission. To test the double morphology of the radio source, we extended the procedure of \cite{Murgia2009} using a double exponential law of the form

\begin{equation}
    I(r) = I_{0,1}e^{-r/r_{\rm e,1}} + I_{0,2}e^{-r/r_{\rm e,2}} ,
    \label{eq:sdouble_exponential}
\end{equation}
where the parameters are the central radio surface brightnesses $I_{0,1},I_{0,2}$ and the corresponding $e$-folding radii $r_{\rm e,1},r_{\rm e,2}$, the length-scales at which the surface brightness drops to $I_0/e$. 
We masked all the discrete sources in the cluster outskirt which would possibly overlap with the extended emission (red regions of \cref{fig:HBA_15asec_and_annuli}, left) and stopped the profile where average surface brightness is 2 times the detection limit in each annulus $\sigma_{\rm annulus} = \sigma_{\rm rms}/\sqrt{N_{\rm beam}}$ (dashed black line in \cref{fig:HBA_15asec_and_annuli}, left), that corresponds to $0.4 R_{500}$ from the cluster centre. The resulting profile is shown in \cref{fig:HBA_15asec_and_annuli} (right) and the best fit parameters are listed in \cref{tab:H_MH_summary}.
We see a clear evidence of an inner brighter component and a second broader emission, with a surface brightness one order of magnitude lower than that of the inner one.
This double-exponential fit allows us to set a boundary on the extension of the inner component. Motivated by the empirical evidence that halos do not extend indefinitely, it is customary to adopt a source radius of $3r_{\rm e}$ \citep{Murgia2009, Botteon2022DR2, Bruno2023_LoTSS-DR2}, which contain 80\% of the flux that would be obtain when integrating the model up to infinity $S_{\nu}(<3r_{\rm e})=0.8S_{\nu}^{\infty}$.
Thus, we fix the size of the mini-halo at three times the $e$-folding radius of the first model component, $R_{\rm MH} = 3r_{\rm e,1} = 84$ kpc. 
The second component results to have a extension of $R_{\rm ext} = 3r_{\rm e,2} = 870$ kpc, which corresponds to a total size $D_{\rm ext} \sim 1.7$ Mpc. This means, double the size based on the $3\sigma_{\rm rms}$-contours of \cref{subsec:radio_morphology}.
We notice that the fitted $e$-folding radius for the inner component is smaller that the value of \cite{Riseley2023}, who measured a total extent of 343 kpc (i.e. a radius of $\sim$ 170 kpc) at 144 MHz, using the same LOFAR observations. 
However, we point out that those authors focus their study on the inner component only, using a higher $3\sigma_{\rm rms}$ threshold and a more robust-weighted imaging parameters for a better spatially-resolved spectral analysis. When using common imaging settings, the results are consistent.
We also notice the presence of several $2\sigma_{\rm rms}$-contours with filamentary morphology, around the central mini-halo of  \cref{fig:HBA_15asec_and_annuli} (left), which we attribute to the extended halo component. 
To exclude the possibility of chance occurrences, we analyse the surface brightness profiles of several other radio sources within the same field of view. 
This examination revealed the absence of plateau-like features  similar to that depicted in \cref{fig:HBA_15asec_and_annuli} (right), confirming its association with the extended radio component.

We then analysed the low resolution ($\mathrm{35''\times35''}$) image, where the broader extended emission can be detected at higher signal to noise (\cref{fig:radio_lowfreq}, right), using Halo-Flux Density CAlculator\footnote{ \href{https://github.com/JortBox/Halo-FDCA}{https://github.com/JortBox/Halo-FDCA}} \citep[\textsc{Halo-FDCA},][]{Boxelaar2021} fitting procedure. 
This method fits the surface brightness profile of halos to a 2D-model using Bayesian Inference and allows to extend the exponential 1D-model, 
taking into account different shapes of halos (circular, elliptical or skewed). Moreover, it is able to perform flux density and size estimation less affected by the image noise. 
The general exponential model used by \cite{Boxelaar2021} is
\begin{equation}\label{eq:FDCA_exponential_profile}
    I(\textbf{r}) = I_0e^{-G(\textbf{r})} ,
\end{equation}
where $I_0$ is the central surface brightness value and $G(\textbf{r})$ the function that takes different forms depending on the complexity of the model. In the circular model case, $G(\textbf{r})=|{\textbf{r}|/r_{\rm e}}$ recalls the exact 2D version of the exponential profile used in \cite{Murgia2009}.
Since \textsc{Halo-FDCA} is not built for a double component emission, we additionally masked the central MH region to derive flux coming from the extended emission only (\cref{fig:FDCA_appendix}). In masked regions, \textsc{Halo-FDCA} performs an extrapolation of the model. The masked circular MH region has a size $R_{\rm MH} =\rm 84\,kpc$ based on the 1D-fitting results.
Despite the clear elliptical shape of the extended radio emission, elongated in the North-South direction (\cref{fig:radio_lowfreq}, right), \textsc{Halo-FDCA} does not find the elliptical model to be a better representation of the source (see \cref{fig:FDCA_appendix} and \cref{tab:H_summary}).
From the circular 2D model, we get the $e$-folding radius $ r_{\rm e}^{\rm FDCA}=224\pm16\, \rm kpc$ (i.e. $R_{\rm ext}^{\rm FDCA} = 3r_{\rm e}^{\rm FDCA}\sim 675$ kpc, which corresponds to $D_{\rm ext}^{\rm FDCA}\sim 1.3$ Mpc) and a flux of $S^{\rm FDCA} = \rm  106\pm8 \,mJy$ integrated until $3r_{\rm e}^{\rm FDCA}$. 
The best-fit results are listed in Tab. \ref{tab:H_summary}. 

The two fitting methods give compatible results for the extended component: $r_{\rm e, 2}=290\pm 60$ kpc and $ r_{\rm e}^{\rm FDCA}=224\pm16\, \rm kpc$, for the 1D and \textsc{Halo-FDCA} $e$-folding radius, respectively.
Moreover, both methods find the second component to be wider compared to the $800\rm\, kpc$ size based on the $\rm 3\sigma_{\rm rms}$ contour of \cref{fig:radio_lowfreq}, with a total linear size $D_{\rm ext}>1.2 \,\rm Mpc$.
This suggests the presence of a second very low surface brightness radio emission filling the cluster volume, overlapped with the previously known central mini-halo.

\begin{table*}
\small
\begin{center}
\renewcommand\arraystretch{1.5}

\caption{\label{tab:H_MH_summary} Summary of the main properties of the radio emission in A1413.}
\vspace{0.4cm}
\begin{tabular}{ c c c c c c c c c  }
\hline
\hline
\multirow{2}{*}{ } & $\nu$       & $D$                    & $S_{3\sigma}$ & $P_{3\sigma}$  & $I_{0,12}$ & $r_{e,12}$ & $\rm \chi$\\ 
                   & \small{MHz} & \small{$\mathrm{kpc}$} & \small{mJy}   & \small{W/Hz}   &  \small{$\rm\mu Jy/arcsec^2$}  &  \small{$\mathrm{kpc}$} & \\ 
\hline
  Halo      & $144$  & $(800,500)$  & $37\pm6$    & $1.9\pm0.3 \,10^{24}$   & $1.6\pm0.4$ & $290\pm60$ & 1.8  \\ 
\hline
  Mini-Halo & $144$  & $220^\star$        & $23\pm2$    & $(1.2\pm0.1)\times 10^{24}$    & $23\pm5$ & $28\pm5$ &   \\ 
            & $1440$ & $220^{ }$        & $1.9\pm0.7$ & $(1.0\pm0.1)\times10^{23}$  & & &   \\ 
 \hline
\end{tabular}
\end{center}
\caption*{
\begin{enumerate*}[start=1,label={Col. \arabic*:},align=left]
    \item Name of the radio source;
    \item size measured above $\rm 3\sigma_{\rm rms}$ contour. For the halo, the two measures refer to $(D_{max},D_{min})$ in the north-south and west-east direction, respectively. Listed result at 1440 MHz from \cite{Govoni2009}. $^{\star}$: Since in LOFAR observations it is impossible to disentangle the mini-halo, we use 220 kpc region as a reference (see \cref{subsec:MH_integrated_spectral_index}).
    \item Flux density measured within the $\rm3\sigma_{\rm rms}$ contour;
    \item Radio power with the $k$-correction applied, assuming $\rm \alpha = -1.2$;
    \item Central surface brightness of the double-exponential profile of \cref{eq:sdouble_exponential};
    \item $e$-folding radius of the double-exponential profile of \cref{eq:sdouble_exponential}; 
    \item reduced $\chi^2$ of the fit.
\end{enumerate*}
}
\end{table*}
\begin{table*}
\small
\begin{center}
\renewcommand\arraystretch{1.5}
\caption{\label{tab:H_summary} Summary of the \textsc{Halo-FDCA} results for the extended emission in A1413.
    }
\vspace{0.4cm}
\setlength{\tabcolsep}{3.5pt}
\begin{tabular}{ l c c c c c c c c c c l}
\hline
\hline
\multirow{2}{*}{ } & $\nu$       & $ I_0 $                      &$ r_{e,1}$     & $r_{e,2}$     & $r_{e,3}$    & $r_{e,4}$   & $S_{<3 r_e}$ & $P_{<3 r_e}$       & $\rm \chi$ &Model & Ref.\\ 
                   & \small{MHz} & \small{$\rm\mu Jy/arcsec^2$} & \small{kpc}   & \small{kpc}   &  \small{kpc} & \small{kpc} & \small{mJy}  &  \small{$\rm W/Hz$} &            &      & \\ 
\hline
  Halo      & $144$  & $2.6\pm0.2$  & $224\pm16$    & -            & -          & -            & $106\pm 8$   & $ (6.0\pm0.5) \times 10^{24}  $          & 0.823  & circle &  \cref{fig:FDCA_appendix} (top panel)\\ 
            &        & $2.6\pm0.2$  & $228\pm17$    & $224\pm16$   & -           & -           & $107\pm 8$   & $ (6.0\pm0.5)\times 10^{24} $  & 0.828 & ellipse & \cref{fig:FDCA_appendix} (middle panel)\\ 
            
            &        & $2.6\pm0.2$  & $295\pm40$    & $227\pm35$   & $148\pm25$  & $255\pm34$  & $109\pm 8$   & $ (6.0\pm0.5)\times 10^{24} $ & 0.769 & skewed & \cref{fig:FDCA_appendix} (bottom panel)\\
\hline
 \hline
\end{tabular}
\end{center}

\caption*{
\begin{enumerate*}[start=1,label={Col. \arabic*:},align=left]
    \item Name of the radio source;
    \item Frequency;
    \item Central surface brightness obtained by \textsc{Halo-FDCA}, \cref{eq:FDCA_exponential_profile};
    \item $e$-folding radius obtained by \textsc{Halo-FDCA} (circular model), 
    \item (elliptical model),
    \item (skewed model)
    \item Flux density measured by \textsc{Halo-FDCA} the $3r_{\rm e}$;
    \item Radio power with the $k$-correction applied, assuming $\rm \alpha = -1.2$;
    \item Reduced $\chi^2$;
    \item Reference image.
    \textcolor{white}{----------------------------------------------------------------------------------------------------------------------------------}
\end{enumerate*}
}

\end{table*}
%
%
%
\begin{figure}
    \centering
    \includegraphics[width=\columnwidth]{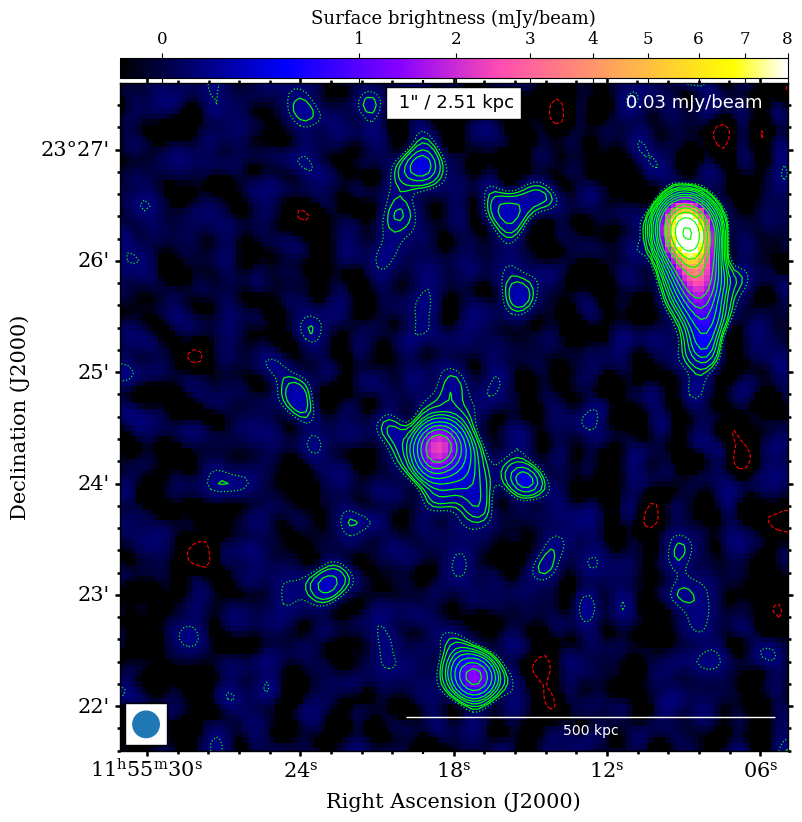}
    \caption{Radio image at 1.5 GHz with beam of $\rm 15\arcsec \times 15\arcsec$.
    The contour levels start at $2\sigma_{\rm rms}$, where  $\mathrm{\sigma_{\rm rms}=0.035 \,m Jy/beam}$, and are spaced with a factor of $\mathrm{\sqrt{2}}$. The $-3\sigma_{\rm rms}$ contours are red coloured, while $2\sigma_{\rm rms}$ contours are dotted green coloured.
    This image, shown here to provide context, is derived and presented in \cite{Govoni2009}.
    }
    \label{fig:VLA_Govoni}
\end{figure}
\begin{figure}[ht]
\centering
  \includegraphics[width=\columnwidth]{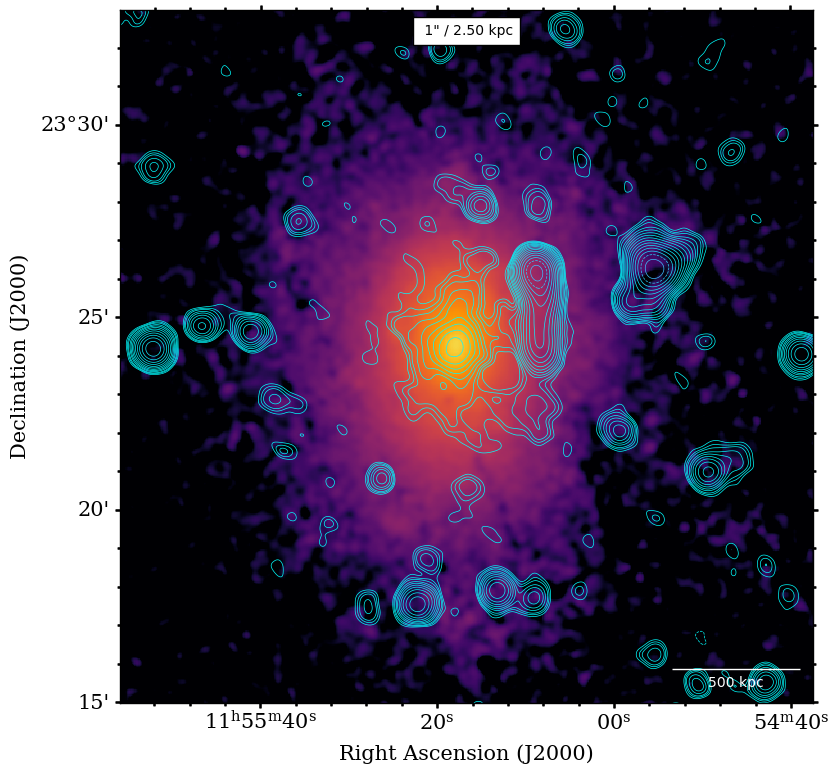}
  \caption{Background subtracted and exposure corrected \XMM image of A1413 in [0.7-2] keV energy band. Radio contours from $35\arcsec$ LOFAR HBA image at 144 MHz \cref{fig:radio_lowfreq} (right) are over plotted.}
     \label{fig:XMM_soft_and_radio_cont}
\end{figure}
%
%
\subsection{Mini-halo integrated spectral index}
\label{subsec:MH_integrated_spectral_index}
The radio mini-halo emission detected by \cite{Govoni2009} (\cref{fig:VLA_Govoni}) is about $1.5'$ ($\simeq$ 220 kpc) in size, and has a flux density of $\mathrm{S_{MH,1.4GHz}=1.9\pm0.7 \, mJy}$, which corresponds to a power of  $\mathrm{P_{MH,1.4GHz}=1.0\cdot10^{23}\, W/Hz}$ (see \cref{tab:H_MH_summary}). 
We caution that this flux density value should be considered as an upper limit to the actual flux density of the MH at 1.5 GHz. Indeed, just the contribution from a single unresolved source (source A in our analysis) detected in the FIRST survey \citep{Becker1995} was considered, whereas an additional point source is embedded within the MH, as shown by our LOFAR high-resolution image (\cref{fig:LR_radio} left, source B).
%
%
To estimate the integrated spectral index of the MH emission, we have re-imaged the LOFAR data at the same $\mathrm{15''\times15''}$ resolution and with the common inner $uv$-cut of $180\lambda$, minimum baseline of VLA image. 
To be consistent with \cite{Govoni2009}, we use a circular region with $\mathrm{r\sim0.7\arcmin\sim110\, kpc}$ and find total flux density of $\mathrm{S_{MH,144MHz} =  23\pm2 \,mJy}$ (Tab. \ref{tab:H_MH_summary}). The integrated spectral index between 144 MHz and 1.5 GHz is $\alpha_{\rm MH}=1.1\pm0.2$. Note that the spectral index might be steeper, considering that the flux reported by \cite{Govoni2009} could be slightly overestimated, as explained above.\\

As the extended halo emission is not detected in the VLA image, we try to put a lower limit to its average spectral index, comparing the mean surface brightness of the extended emission at 144 MHz with the rms noise of VLA image. To do this, we consider a rectangular area that follows the $3\sigma_{\rm rms}$ contours of the 800 kpc scale emission (\cref{fig:radio_lowfreq}, right) and exclude the central $\mathrm{r<0.7'}$. 
Unfortunately, this does not allow us to firmly constrain the spectral index of the external emission, as we find $\alpha>0.75$ at $1\sigma_{\rm rms}$. \cite{Riseley2023} produced resolved spectral map combining MeerKAT at 1283 MHz with LOFAR at 145 MHz data and found that the innermost region ($R \lesssim 50$ kpc, which corresponds to $\sim2r_{\rm e,1}$ of our inner component) shows a flatter spectral index around $−0.7$ to $−1$, while there is a hint of larger-scale diffuse emission with a typical spectral index of around $−1.1$ or steeper. The different spectral index of the two regions may suggest the presence of two different types of radio emission co-existing in the same cluster. 
Further investigation about the hybrid morphology of the radio emission is performed in  \cref{subsec:spatialcorrelation}, where also X-ray properties are considered.\\

\section{Discussion}\label{sec:discussion}

\subsection{Spatial correlation between X-ray and radio surface brightness}\label{subsec:spatialcorrelation}
Quantitative studies to investigate the point-to-point correlation between the radio and X-ray brightness distribution of giant halos can provide useful information about the connection between the thermal and non-thermal emission \citep[e.g.][]{Govoni2001,Feretti2001,Giacintucci2005}. However, most of the literature studies were limited by the resolution and sensitivity of the radio and X-ray observations. 
Now, thanks to deep and sensitive observations, these studies can be carried out in more detail and extending the sample also to mini-halos \citep{Ignesti2020,Botteon2020,Rajpurohit2021,Biava2021,Duchesne2021, Bonafede2022,Riseley2022a,Riseley2022b}. 
These recent studies have shown different trends for giant and mini-halos: halos turned out to show a linear or sub-linear scaling, while mini-halos tend to have a super-linear one.

The correlation is generally investigated via a power-law relationship of the kind:
\begin{equation}
    \log{I_R} = A\,\log{I_X}+B ,
    \label{eq:Ir_Ix_relation}
\end{equation}
where the slope of the scaling $A$ determines whether the radio brightness (i.e. the magnetic field strength and CRe density) declines faster ($A > 1$) than the X-ray brightness (i.e. the thermal gas density), or vice versa ($A < 1$). 
The connection between thermal and non-thermal plasma could reflect the particle re-acceleration mechanism, and it could be used to discriminate between models of halo formation \citep{Govoni2001}.

%
%
%
\begin{table}[t]
\begin{center}
\renewcommand\arraystretch{1.6}
\caption{\label{tab:LinMix_result}\texttt{\footnotesize LinMix} best fit parameter for $I_R-I_X$ correlation of \cref{fig:correlations}.}
\vspace{0.4cm}
\setlength{\tabcolsep}{3.5pt}
\begin{tabular}{ lccccl }
\hline
\hline
 Region & Grid & $A$ & $r_p$  & $r_S$ & Ref. \\ 
\hline
 Halo & $35'' \times35''$ & $0.81^{+0.07}_{-0.08}$ & $0.87$ & $0.85$ & \cref{fig:correlations} (left)\\
      & $35'' \times35''$ & $0.78^{+0.08}_{-0.07}$ & $0.83$ & $0.82$ & \cref{fig:correlations} (left)\\
\hline
%
 Mini-halo & $23'' \times23''$ & $1.27^{+0.16}_{-0.14}$ & $0.93$ & $0.91$ & \cref{fig:correlations} (right)\\
 \hline
\end{tabular}
\end{center}
\caption*{
\begin{enumerate*}[start=1,label={Col. \arabic*:},align=left]
    \item Considered radio source;
    \item Grid boxes size;
    \item fitting slopes ($A$);
    \item Pearson ($r_p$), and
    \item Spearmann ($r_s$) correlation coefficients; 
    \item Reference plot. \textcolor{white}{--------------}   
\end{enumerate*}
}
\end{table}

To compute these correlations we used the \texttt{\footnotesize LinMix}\footnote{\href{https://linmix.readthedocs.io/en/latest/index.html}{https://linmix.readthedocs.io/en/latest/index.html}} package \citep{Kelly2007}, which uses a Bayesian approach to linear regression and allows to consider measurement uncertainties on both quantities. 
The threshold for upper-limits is fixed at $2\sigma_{\rm rms}$ in the radio image following \cite{Botteon2020}.
These authors noted that the introduction of a high threshold combined with a large intrinsic scatter can introduce a bias in the correlation and demonstrated that threshold of $2\sigma_{\rm rms}$ is the good choice.
The strength of the correlation is measured by the Pearson $r_p$ and Spearmann $r_s$ correlation coefficients. The former assessing the linear relation between the two variables, while the latter determining their monotonic relation (whether linear or not).

Based on the two-component emission found with the surface brightness analysis of \cref{subsubsec:radio_SB}, we performed point-to-point correlation separately for both the extended and mini-halo emission, using the radio images at 144 MHz presented in Fig. \ref{fig:radio_lowfreq} (right) and \cref{fig:HBA_15asec_and_annuli} (left) together with the XMM-Newton images presented in Fig. \ref{fig:XMM_soft_and_radio_cont}.
Following \cite{Govoni2001}, we constructed a grid covering the cluster region and calculated for each box the mean radio and X-ray quantities, as well as the root-mean-square (rms), which can be assumed as an estimate of the statistical error. 
The size of the boxes has been chosen to be equal to the 1.5 times the area of the radio beam (see \cref{tab:LinMix_result}). This grid resolution also allows to be minimally affected by the PSF of XMM-Newton, which for the combined MOS+pn images is $\sim 10''$ (FWHM).

\begin{figure*}[ht]
 \centering
 \includegraphics[width=\textwidth, height=7cm]{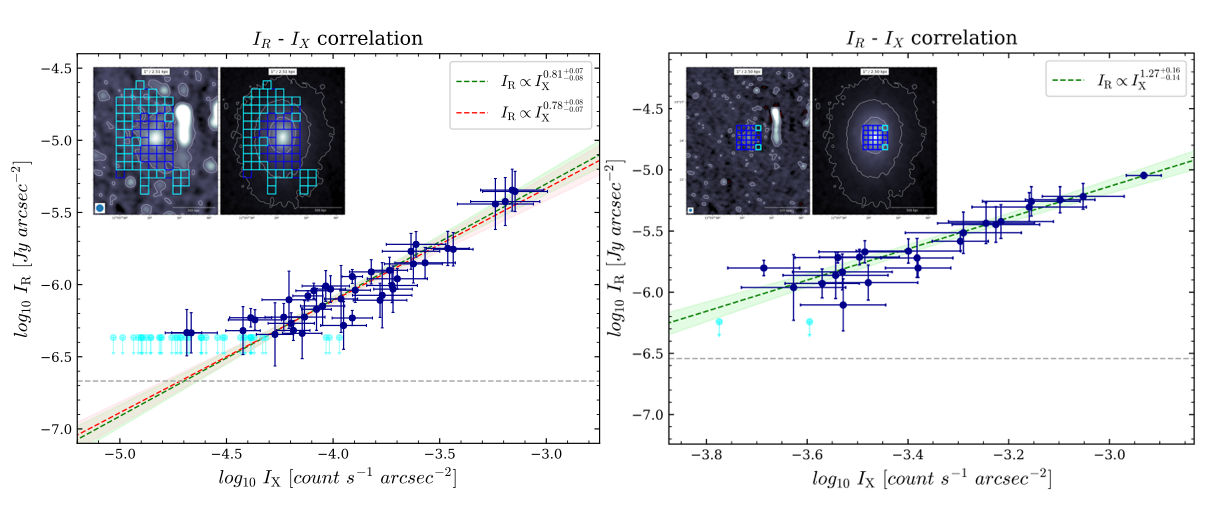}
 \caption{Spatial correlation between X-ray and radio surface brightness.
 Left panel: $I_R-I_X$ relation of the halo in A1413, extracted in square boxes with width of $35''$. The horizontal gray dash-dotted lines indicate the $\sigma_{\rm rms}=298\, \mu Jy/$beam in the radio maps. 
 Right panel: $I_R-I_X$ relation of the mini-halo in A1413, extracted in square boxes with width of $23''$. The horizontal gray dash-dotted lines indicate the $\sigma_{\rm rms}=203\,\mu Jy/$beam in the radio maps. 
 For both relations, upper limits (cyan points) refer to cells where the radio surface brightness is below the $2\sigma_{\rm rms}$ level. The best-fit (green dashed line) is reported with the corresponding $1\sigma$ confidence interval. Red dashed line best-fit line (left panel) refers to the correlation without the 4 central boxes.
 }
 \label{fig:correlations}
\end{figure*}

First, we performed the correlation on the 35$\arcsec$-resolution image (\cref{fig:radio_lowfreq}, right) to study the properties of the extended emission.
The total considered area was chosen based on the lowest X-ray brightness level, that includes the radio emission and excludes other external radio sources, in order to select an homogeneous and not radio-biased area. 
Both point-like and extended (the tailed galaxy) contaminating sources in the radio image were identified and cells containing them, removed. Presumably, the tailed galaxy also influences some contiguous regions south of the tail, thus, a safe choice is to exclude these regions too. Final selection (blue boxes) and upper limits (cyan boxes) are shown in the top left corner of each plot of \cref{fig:correlations}.
%
%
%
%
%
%
%
The two surface brightnesses are positively correlated $r_p=0.87, r_s=0.85$ and have a sub-linear scaling $A=0.81^{+0.07}_{-0.08}$. This is in agreement with other literature works for giant halos \citep[e.g.][]{Botteon2020,Rajpurohit2021,Riseley2022a}.
We notice four points, spatially corresponding to the central brighter component, that appear distinctly separated from the extended component ones on the correlation. To exclude that those points are biasing the correlation towards steeper values, we performed it again, removing these four central boxes. 
This also allows to study the correlation properties of the outer component only. We still find a sub-linear ($A=0.78^{+0.08}_{-0.07}$)) strong ($r_p=0.83, r_s=0.82$) correlation, consistent with the previous one. The absence of statistical difference between the two results confirms that the correlation is determined by the extended emission and reinforces the robustness of our initial findings.

The correlation indicates that the decline of the non-thermal radio component is shallower than the thermal one: this slope can be explained by re-acceleration models, while it is a challenge for hadronic models \citep[as explained by][]{Govoni2001}.

We then study the central mini-halo component, making use of the higher 15$\arcsec$-resolution radio image (\cref{fig:HBA_15asec_and_annuli}), that is the most sensitive to the central emission and allows to better sample the inner region.
Differently from the sub-linear or linear scalings that are reported in the literature for giant radio halos, but consistently with the sample of MHs studied by \cite{Ignesti2020}, we find instead a super-linear $A=1.27^{+0.16}_{-0.14}$ correlation for the central emission (\cref{fig:correlations}, right). Also in this case, there is a clear evidence of a spatial correlation between radio and X-ray emission, with $r_p=0.93, r_s=0.91$ (\cref{tab:LinMix_result}).
This super-linear correlation is also consistent with analysis of \cite{Riseley2023}, who found $A=1.20^{+0.13}_{-0.11}$ using LOFAR data at 144 MHz and $A=1.63^{+0.10}_{-0.10}$ reproducing the correlation with MeerKAT data at 1283 MHz. 
The super-linear scaling between $I_R$ and $I_X$ suggests that the number density of emitting electrons and magnetic field rapidly declines from the center to the external regions. In particular, the radio emission is more peaked than the thermal emission, indicating that the ICM non-thermal component is more concentrated around the central AGN. This is consistent with the hadronic scenario that predicts a super-linear scaling between the surface brightness for a radially decreasing magnetic field \citep[e.g.][]{Govoni2001}. 
We note, however, that the re-acceleration model cannot be excluded. Depending on the CRe distribution, it could also produce a super-linear scaling of the X-ray and radio surface brightness.

The different trends of the two radio sources further strengthens the idea that the radio emission of A1413 is composed by two different components. They might also mean that the mechanisms responsible for the re-acceleration of the radio emitting particle are different.


\section{Conclusions}
\label{sec:conclusion}
In this work, we have presented a composite X-ray and radio analysis of the galaxy cluster A1413. For this purpose, \XMM and LOFAR 144 MHz data were first analysed separately. They were then combined to quantify the connection between the thermal and non-thermal components of the ICM. The main results can be summarised as follows:

\begin{enumerate}
  \item The X-ray morphology of the cluster on large scales is not spherically symmetric, confirming the ellipticity found in previous work in, both, optical and X-ray bands. This morphology suggests the occurrence of a merger that has left a clear, though minor, imprint on the gas distribution along the N-S axis of the cluster.
  Other dynamical features indicate that A1413 is an intermediate phase cluster:
  The temperature profile in the core does not decline sharply as it is expected in the inner regions of a strong cool-core cluster. Moreover, the metallicity profile is flatter than expected for a very relaxed system. However, the surface brightness profile is well-fitted with a double $\beta$-model, which typically accounts for the central brightness excess of cool-core clusters.
  Finally, from the calculation of the cooling time, we obtain $t_{\rm cool} = 6.4\pm1.3\,\rm Gyr$ within $r<0.5\arcmin\sim75\,\rm kpc$, and a cooling region $r_{\rm cool} \sim 90\,\rm kpc$.
  We argue that such a system with a moderate cooling time, an elevated central entropy and a slightly decreasing central temperature profile can be classified as a weak cool-core system \citep{Hudson2010}.
  \item A1413 was known to host a mini-halo (\cite{Govoni2009, Giacintucci2017, Savini2019, Riseley2023}).
  The LOFAR HBA data at 144 MHz led to the discovery of an extended low surface brightness emission, with an LLS of about 800 kpc in the N-S direction based on the $3\sigma_{\rm rms}$-contours, so far detected only at low frequencies. 
  To date, this is one of the few objects that show a wider emission surrounding a previously known central mini-halo and with a size typical of giant radio halos. 
  \item  We perform 1D surface brightness fitting using the double exponential model of \cref{eq:sdouble_exponential}. We find an inner brighter component extended up to $R_{\rm MH}=84$~kpc, which is consistent with other mini-halos, and a broader component with $R_{\rm ext}=870$~kpc.
  Additionally, we performed a 2D fitting of the extended component only using the \textsc{Halo-FDCA} software. To do that, we masked the internal region, and we obtained best fitting values of $R_{\rm ext}^{\tiny FDCA} = 3r_{\rm e}^{\tiny FDCA}\sim 675$ kpc for the circular model (see \cref{tab:H_summary}). 
  Both methods give consistent results and find the second component to be wider than the $800\rm\, kpc$ size based on the $\rm 3\sigma_{\rm rms}$ contour of \cref{fig:radio_lowfreq}, with a total linear size $D_{\rm ext}>1.2 \,\rm Mpc$.
  \item We performed a point-to-point correlation analysis between the X-ray and radio surface brightness, which confirms the spatial connection between the non-thermal and thermal emission in galaxy clusters. We investigated a possible correlation for, both, the halo and the mini-halo regions separately.
  We found a sub-linear correlation $I_R \propto I_X^{0.78^{+0.08}_{-0.07}}$
  when analysing the extended part and a super-linear $I_R \propto I_X^{1.27^{+0.16}_{-0.14}}$ correlation for the inner mini-halo component. Sub-linear or linear scalings are reported in the literature for giant radio halos, and can be explained by re-acceleration scenario. On the other hand, super-linear correlations are usually found for mini-halos, indicating that the ICM non-thermal component is more concentrated around the central AGN. This is consistent with the hadronic scenario. Nevertheless, re-acceleration model cannot be excluded, as depending on the CRe distribution, it could also produce a super-linear scaling.

\end{enumerate}

The properties of this cluster show all the signs of an intermediate-state cluster, dynamically not fully relaxed and with diffuse radio emission outside the central cooling region, as would be expected for cool-core objects.
Both the SB fitting and the $I_R-I_X$ correlations confirm the presence of two components as found in other clusters with the same dynamical characteristics (e.g., RX
J1720.1$+$2638, \cite{Biava2021}; A2142, \cite{Bruno2023_A2142}).
The mini-halo and the diffuse emission extend over different scales and show different features, suggesting that the mechanisms responsible for the re-acceleration of the radio-emitting particles are different.
The outer component might probe turbulent re-acceleration induced by a less energetic merger events. In this context, we argue that minor mergers events can induce particle acceleration on large scales without fully disrupting the cluster core.
Considering its hybrid radio morphology and intermediate X-ray properties, A1413 is a peculiar object that challenges the long lasting picture of disturbed galaxy clusters hosting giant radio halos and relaxed ones having mini-halos.


\begin{acknowledgements}
This research project made use of the following Python packages: APLpy \citep{APLpy2012}, Astropy \citep{Astropy2013} and NumPy \citep{Numpy2011}. 
CJR and ABonafede acknowledge financial support from the ERC Starting Grant `DRANOEL', number 714245. 
LL acknowledges financial contribution from the INAF grant 1.05.12.04.01. 
MB acknowledges support from the Deutsche Forschungsgemeinschaft under Germany's Excellence Strategy - EXC 2121 "Quantum Universe" - 390833306 and from the BMBF ErUM-Pro grant 05A2023.
RJvW acknowledges support from the ERC Starting Grant ClusterWeb 804208.
ABotteon acknowledges financial support from the European Union - Next Generation EU.
LOFAR \citep{vanHaarlem2013} is the Low Frequency Array designed and constructed by ASTRON. It has observing, data processing, and data storage facilities in several countries, which are owned by various parties (each with their own funding sources), and that are collectively operated by the ILT foundation under a joint scientific policy. The ILT resources have benefited from the following recent major funding sources: CNRS-INSU, Observatoire de Paris and Université d'Orléans, France; BMBF, MIWF-NRW, MPG, Germany; Science Foundation Ireland (SFI), Department of Business, Enterprise and Innovation (DBEI), Ireland; NWO, The Netherlands; The Science and Technology Facilities Council, UK; Ministry of Science and Higher Education, Poland; The Istituto Nazionale di Astrofisica (INAF), Italy.
This research made use of the Dutch national e-infrastructure with support of the SURF Cooperative (e-infra 180169) and the LOFAR e-infra group. The Jülich LOFAR Long Term Archive and the German LOFAR network are both coordinated and operated by the Jülich Supercomputing Centre (JSC), and computing resources on the supercomputer JUWELS at JSC were provided by the Gauss Centre for Supercomputing e.V. (grant CHTB00) through the John von Neumann Institute for Computing (NIC).
This research made use of the University of Hertfordshire high-performance computing facility and the LOFAR-UK computing facility located at the University of Hertfordshire and supported by STFC [ST/P000096/1], and of the Italian LOFAR IT computing infrastructure supported and operated by INAF, and by the Physics Department of Turin university (under an agreement with Consorzio Interuniversitario per la Fisica Spaziale) at the C3S Supercomputing Centre, Italy.
\end{acknowledgements}

%
%

\bibliographystyle{aa}
\bibliography{mybib}



\begin{appendix}
\onecolumn
\section{Flux Density CAlculator results}
\label{sec:appendix_FDCA}
\begin{figure*}[h!]
    \centering
    \begin{subfigure}{\textwidth}
        \includegraphics[width=\textwidth]{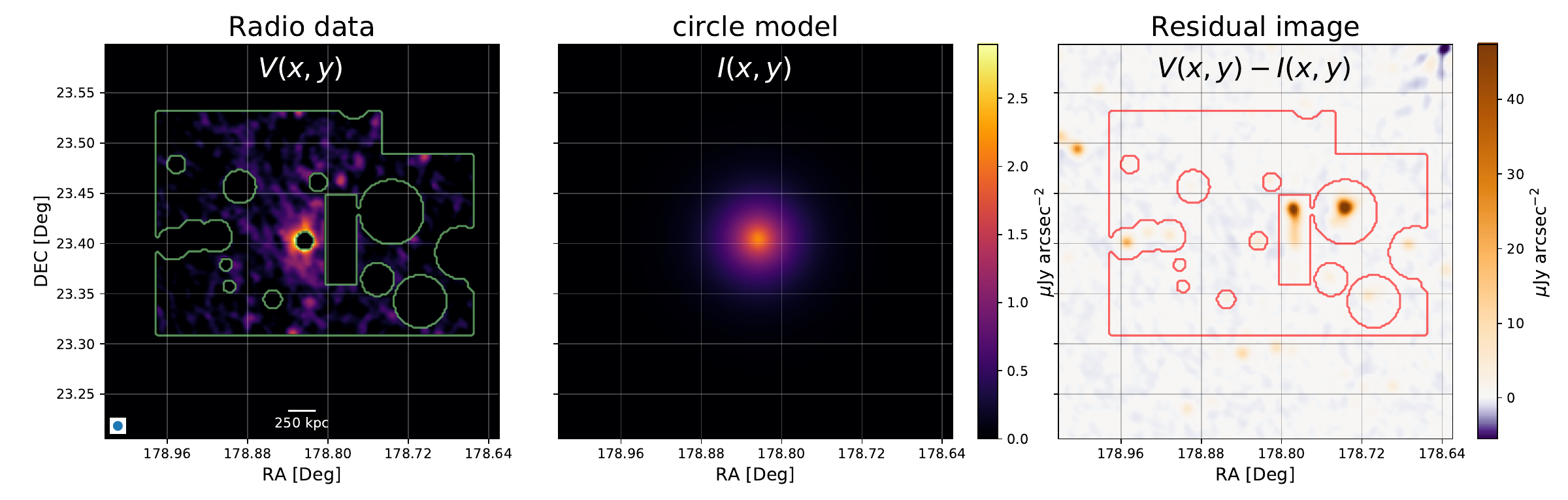}
    \end{subfigure}
    \begin{subfigure}{\textwidth}
        \includegraphics[width=\textwidth]{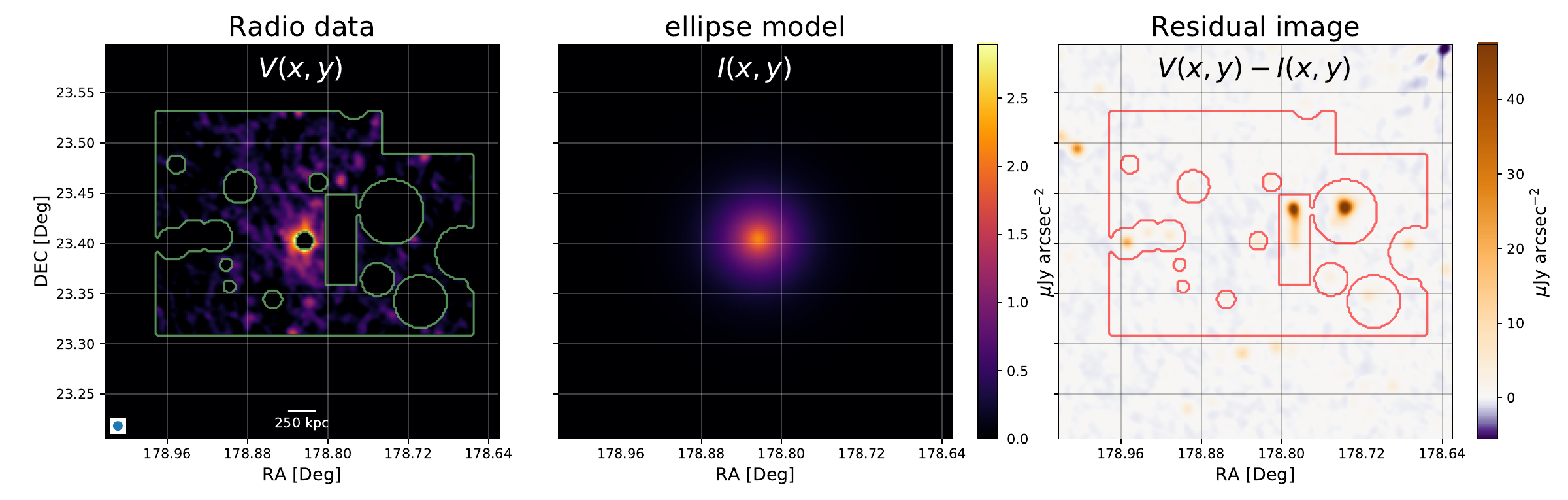}
    \end{subfigure}
    \begin{subfigure}{\textwidth}
        \includegraphics[width=\textwidth]{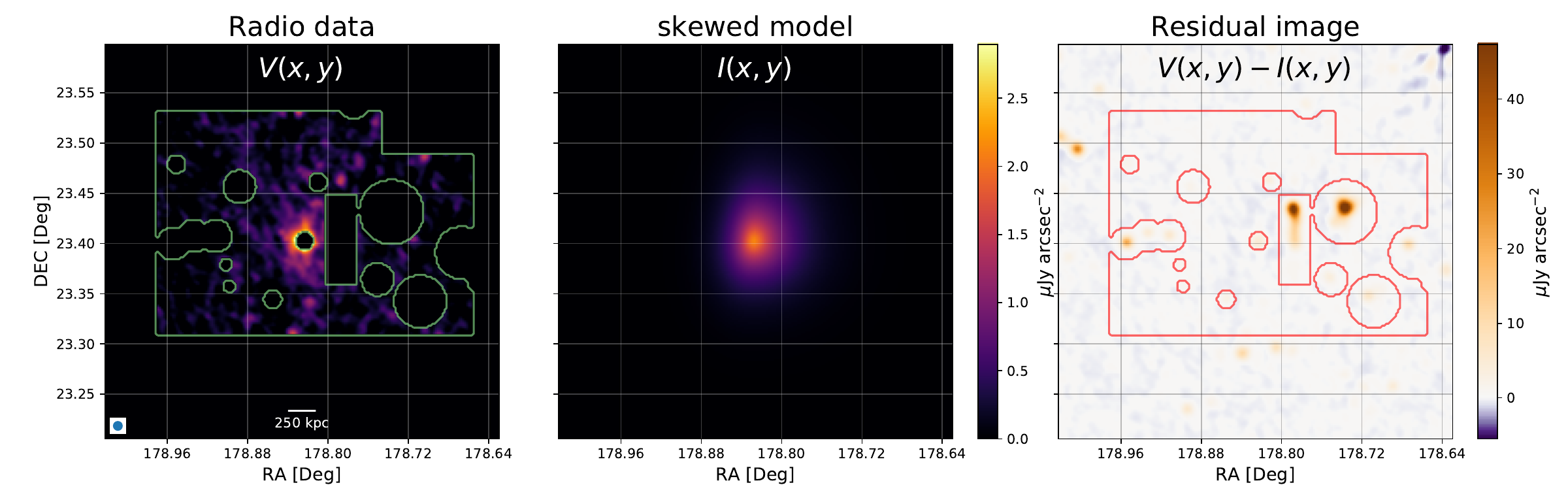}
    \end{subfigure}
    \caption{\texttt{FDCA-Halo} results for the radio halo in A1413 at 144 MHz frequency.
    Left panel: Original image with the contaminating regions masked out. Middle panel: Circular model map. Right panel: Residual image. The red contours show the masked regions, the contamination sources are visible.
    }
    \label{fig:FDCA_appendix}
\end{figure*}
\end{appendix}

\end{document}